\documentclass[pdftex,11pt,a4paper,parskip=half,headsepline]{scrartcl}

\usepackage[english]{babel}
\usepackage[a4paper,left=3cm,right=3cm,bottom=3.5cm,top=3cm]{geometry}
\usepackage[onehalfspacing]{setspace}
\usepackage[automark]{scrlayer-scrpage}
\usepackage[utf8]{inputenc}
\usepackage{latexsym}
\usepackage[T1]{fontenc}
\usepackage{amsmath}
\usepackage{amssymb}
\usepackage{amsfonts}
\usepackage{amstext}
\usepackage{dsfont}
\usepackage{amsthm}
\usepackage{extarrows}
\usepackage{graphicx}
\usepackage{xcolor}
\usepackage{longtable}
\usepackage[longtable]{multirow} 
\usepackage{booktabs} 
\usepackage{listings}
\usepackage{url}
\usepackage{hyperref}
\usepackage{todonotes}

\usepackage[backend=biber, bibstyle=apa, maxcitenames=2, citestyle=apa, isbn=false, doi=false, url=false, eprint=true]{biblatex}
\AtEveryBibitem{\clearfield{month}}
\AtEveryCitekey{\clearfield{month}}
\AtEveryBibitem{\clearfield{day}}
\AtEveryCitekey{\clearfield{day}}
\AtEveryBibitem{\clearfield{note}}
\AtEveryCitekey{\clearfield{note}}
\NewBibliographyString{toappear}
\DefineBibliographyStrings{english}{%
  toappear = {to appear in:},
  version = {R package version},
} 
\usepackage{csquotes}

\addto\extrasenglish{

}

\definecolor{TUGreen}{rgb}{0.517,0.721,0.094}
 \definecolor{middlegray}{rgb}{0.5,0.5,0.5}
 \definecolor{lightgray}{rgb}{0.8,0.8,0.8}
 \definecolor{orange}{HTML}{e88909}
 \definecolor{yac}{rgb}{0.6,0.6,0.1}
 \definecolor{rred}{rgb}{0.8500,0.3250,0.0980}
 \definecolor{turquoise}{HTML}{47C2D6}
 \definecolor{ggreen}{HTML}{1bd38a}
 \definecolor{violet}{HTML}{af93e7}
 \definecolor{red}{HTML}{ea7360}
 \definecolor{pink}{HTML}{f3106d}
 \definecolor{lgreen}{HTML}{94e017 }

\author{Marléne Baumeister$^{*,1,2}$, Konstantin Emil Thiel$^{1,3}$, Lynn Matits$^{4,5}$, \\
 Georg Zimmermann$^{3,6,7}$, Markus Pauly$^{1,2}$, Paavo Sattler$^{1}$}
\title{Multivariate and Multiple Contrast Testing in General Covariate-adjusted Factorial Designs}

\subtitle{Short title: Multiple Contrast Testing in MANCOVA}

\makeatletter

\clearpairofpagestyles
\ihead{\headmark}
\automark{section}
\ohead{Multiple Contrast Testing in MANCOVA}
\newtheorem{satz}{Theorem}[]

\newtheorem{prop}[satz]{Proposition}

\newcommand{\br}{\mathbb{R}}

\DeclareMathOperator{\prob}{P}
\DeclareMathOperator{\E}{E}
\DeclareMathOperator{\var}{Var}
\DeclareMathOperator{\cov}{Cov}

\begin{document}

\thispagestyle{empty}
\maketitle

\renewcommand*{\thefootnote}{\fnsymbol{footnote}}
\footnotetext[1]{Corresponding author: \textsf{e-mail: baumeister@statistik.tu-dortmund.de}}
\renewcommand*{\thefootnote}{\arabic{footnote}}
\footnotetext[1]{Department of Statistics, TU Dortmund University, Germany}
\footnotetext[2]{Research Center Trustworthy Data Science and Security, UA Ruhr, Germany}
\footnotetext[3]{Research Program Biomedical Data Science, Paracelsus Medical University Salzburg, Austria.}
\footnotetext[4]{Clinical \& Biological Psychology, Institute of Psychology and Education, Ulm University, Ulm, Germany}
\footnotetext[5]{Sports and Rehabilitation Medicine, Department of Medicine, Ulm University Hospital, Ulm, Germany}
\footnotetext[6]{Team Biostatistics and Big Medical Data, IDA Lab Salzburg, Paracelsus Medical University, Salzburg, Austria}
\footnotetext[7]{Department of Artificial Intelligence and Human Interfaces, Paris Lodron University, Salzburg, Austria}

\section*{Abstract}
Evaluating intervention effects on multiple outcomes is a central research goal in a wide range of quantitative sciences. 
It is thereby common to compare interventions among each other and with a control across several, potentially highly correlated, outcome variables. 
In this context, researchers are interested in identifying effects at both, the global level 
(across all outcome variables) and the local level (for specific variables).
At the same time, potential confounding must be accounted for. 
This leads to the need for powerful multiple contrast testing procedures (MCTPs) capable of handling multivariate outcomes and covariates.
Given this background, we propose an extension of MCTPs within a semiparametric MANCOVA framework that allows applicability beyond multivariate normality, homoscedasticity, or non-singular covariance structures.
We illustrate our approach by analysing multivariate psychological intervention data, evaluating joint physiological and psychological constructs such as heart rate variability. 
 
\textbf{Keywords:} Multiple Testing, Covariate Adjustment, Multivariate Factorial Design, Interventional Study, Bootstrap.

\section{Introduction}\label{sec:intro}
In biological, medical, and psychological research, there is a strong need for multiple testing methods, as such questions 
often arise in factorial design that are common for these fields. 
Typically, post hoc tests are considered after rejecting a global hypothesis when comparing more than two groups. 
This occurs, e.g., in clinical studies, or in ecological studies. 
An example of the latter is the comparison of the relative reproductive success (fitness) of birds grouped by sex and colour morph \parencite{boerner_aggression_2009}.
On the other hand, in psychological intervention studies, there are often only two groups (intervention and control), but multiple, often highly correlated outcomes are measured. 
Indeed, correlated outcomes are inherent in the structure of intervention studies: in particular, if several related measures are taken from the same individual at a single time point, it is natural to assume a (strong) dependency between these outcome variables. 
In such settings, joint modelling of these outcomes is more suitable, and statistical testing 
usually becomes more reliable when it accounts for the underlying dependencies.
\textcite{warne_primer_2014} explicitly recommends using a multivariate model in such situations to avoid type I error inflation.

To exemplify the practical issues, we consider a synthetic dataset \parencite{thiel_hypnotreatsynth_2025} based on original data from the intervention-based \textit{HypnoTreat} study conducted at the University of Ulm \parencite{karrasch_effects_2022,karrasch_exploratory_2023,karrasch_randomized_2023}.
The study examined the effects of a single relaxation hypnosis session on psychological and biological variables in chronically stressed individuals, such as heart rate variability.
Here, the intervention-induced changes are observed in multiple variables measuring the same physiological construct.
Consequently, we are interested in an overall global effect ("Does hypnosis influence HRV?") and specific local effects ("Does hypnosis influence a single specific parameter?"). 
This requires testing of multiple hypotheses. 
Moreover, modelling and inferring these questions jointly in one model, requires a multivariate approach. 
Additionally, the data set contains confounding covariates such as perceived chronic stress and suggestibility (ability to be hypnotised), which have to be accounted for. 
In fact, including covariates in a statistical analysis usually leads to an increase in power as
predictive covariates can explain variability in the outcomes, which improves detection of factorial effects \parencite{thiel_resampling_2024}. 
Therefore, it is recommended to adjust for covariates if they are of predictive character \parencite{kahan_risks_2014}.
In line with this, also regulatory authorities \parencite{ema_guideline_2015, us_department_adjusting_2023} recommend covariate-adjustment in randomized clinical trials.

The co-occurences of multivariate outcomes, predictive covariates, and multiple testing problems, motivate the adaption of multiple contrast testing procedures (MCTPs) to a semiparametric MANCOVA framework. 
MCTPs are a powerful solution for multiple testing as they redefine the rejection of the global hypothesis: the global hypothesis is rejected simultaneously if one of the local hypotheses is rejected.
By construction, it is transparent which of the local tests are responsible for the global rejection and both local and global test decisions are coherent and consonant \parencite{bretz_multiple_2011}.
Traditional approaches, such as ANOVA followed by adjusted post hoc t-tests, do not have these advantages.
\textcite{konietschke_are_2013} pointed out that MCTPs provides more information about the process of rejecting the hypotheses.
This is because the testing principle of MCTPs complies with the union-intersection principle introduced by \textcite{roy_heuristic_1953}.
Moreover, as MCTPs are simultaneous testing procedures, they allow for the construction of compatible confidence intervals.

Furthermore, in many situations MCTPs turn out to be more powerful than methods of classical p-value adjustment like the Bonferroni procedure \parencite{bretz_multiple_2011}. 
Because of this, MCTPs are known to be effective for many models and estimands.
There are different methods for various univariate 
\parencite{bretz_numerical_2001, hasler_multiple_2008, konietschke_are_2013, baumeister_early_2025}, 
multivariate \parencite{hasler_multiple_2014, sattler_quadratic_2024} and even high dimensional and functional scenarios with metric outcomes 
\parencite{konietschke_small_2021, munko_general_2023}. 
Moreover, there are even rank-based MCTPs for univariate outcomes \parencite{konietschke_rank-based_2012, noguchi_nonparametric_2020}, repeated measures \parencite{umlauft_wild_2019}, and other complex designs \parencite{rubarth_estimation_2022}. 
\textcite{hasler_dunnett-type_2011} and \textcite{hasler_multiple_2014} where the first to introduce MCTPs for multiple endpoints, but they did not allow for covariate adjustment and assume a parametric model. 
More recently, \textcite{becher_analysis_2025} introduced covariate-adjusted MCTPs, but only allow for one univariate outcome. 
To reach our goal, we consider a more general semiparametric MANCOVA model, as studied in \textcite{zimmermann_multivariate_2020}. 
The method allows for global testing in factorial designs and considers a covariate-adjusted mean as estimand. 
As it allows for covariance heteroscedasticity and some types of singularity, the model is very flexible. 
However, multiple testing was not considered so far. 
We close this gap, by proposing a MCTP for this general framework. 
In particular, our contribution is an MCTP for multivariate covariate-adjusted means
\begin{enumerate}
\item[i.]  that is asymptotically valid in general semiparametric models without assuming normality, while
\item[ii.] allowing for potential covariance heteroscedasticity and singularity,
\item[iii.] that covers various local and global testing problems such as multivariate Dunnett- or 
Tukey-type hypotheses, especially in intervention designs.
\end{enumerate}
To improve small sample performance, we consider two resampling approaches: parametric and wild bootstrapping.
Both have demonstrated accurate performances, not only in \textcite{zimmermann_multivariate_2020}, but also in other multivariate factorial designs \parencite{konietschke_parametric_2015, friedrich_wild_2017, friedrich_mats_2018}, and also in multiple testing \parencite{umlauft_wild_2019, munko_general_2023, munko_rmst-based_2024}.

The paper is structured as follows. 
In Section \ref{sec:set-up} we present the semiparametric MANCOVA model and state central limit theorems. 
All statistical methods are given in Section \ref{sec:methods}, including singularity-robust covariance estimator, the general multiple testing problem, and specific cases for concrete testing problems.
In addition, we state and explain asymptotic guarantees for the bootstrap, the determination of proper critical values, and the resulting MCTP including local p-values. 
In Section~\ref{sec:simus}, we evaluate the method’s small-sample properties via extensive simulation. 
We thereby examine family-wise type I error rate (FWER) control and power of the MCTP in comparison with existing methods. 
Then, we present an illustrative data analysis based on the \textit{HypnoTreat} study in Section \ref{sec:dataanalysis}.
The paper closes with a discussion in Section \ref{sec:end}.

\section{Statistical Model and Set-Up}\label{sec:set-up}
We consider a general MANCOVA set-up with $d$-dimensional random variables $\mathbf{Y}_{ij}=(Y_{ij1},\dots,$ $Y_{ijd})'$ representing the outcome of individual $j\in\{1,\dots,n_i\}$ in group $i\in \{1,\dots,k\}$. 
Attached to each outcome vector there is a $c$-dimensional individual-specific covariate vector $\mathbf{z}_{ij}=(z_{ij1},\dots,z_{ijc})'$  and we pool all outcome and covariate vectors in the $n$-dimensional vector $\mathbf{Y}$ $=(\mathbf{Y}_{11}',\dots,\mathbf{Y}_{kn_k}')'$ and the  $n\times c$ matrix $\mathbf{Z}=(\mathbf{z}_{11},\dots,\mathbf{z}_{kn_k})'$, where $n:=\sum_{i=1}^kn_i$.
In the following, let $\mathbf{I}_d$ denote the $d$-dimensional identity matrix, $\mathbf{1}_d$ a $d$-dimensional vector and $\mathbf{J}_d$ a $d$-dimensional quadratic matrix containing only 1s.
Moreover, $\oplus$ denotes the direct sum and $\otimes$ denotes the Kronecker product of matrices.
Additionally, we introduce a vector of $n$ error variables $\boldsymbol{\epsilon}=(\boldsymbol{\epsilon}_ {11}',\dots, \boldsymbol{\epsilon}_{kn_k}')'$, $\boldsymbol{\epsilon}_{ij}=(\epsilon_{ij1},\dots,\epsilon_{ijd})'$, a vector of $k$ adjusted means $\boldsymbol{\mu}=(\boldsymbol{\mu}_ {1}',\dots, \boldsymbol{\mu}_{k}')'$, $\boldsymbol{\mu}_{i}=(\mu_{i1},\dots,\mu_{id})'$ and a vector of $c$ regression coefficients $\boldsymbol{\nu}=(\boldsymbol{\nu}_ {1}',\dots, \boldsymbol{\nu}_{c}')'$, $\boldsymbol{\nu}_{m}=(\nu_{m1},\dots,\nu_{md})',\,i\in\{1,\dots,k\},\,j\in\{1,\dots,n_i\},\,m\in\{1,\dots,c\}$. 
Then our general semiparametric MANCOVA model is given by
\begin{align*}
\mathbf{Y}=\tilde{\mathbf{M}}\boldsymbol{\mu}+\tilde{\mathbf{Z}}\boldsymbol{\nu}+\boldsymbol{\epsilon},
\end{align*}
where $\tilde{\mathbf{M}}=\bigoplus_{i=1}^k(\mathbf{1}_{n_i}\otimes\mathbf{I}_d)$ and $\tilde{\mathbf{Z}}=\mathbf{Z}\otimes\mathbf{I}_d$.
Thereby, $\tilde{\mathbf{X}}=(\tilde{\mathbf{M}}, \tilde{\mathbf{Z}})$ is the \emph{design matrix} of a linear model, where $\tilde{\mathbf{M}}$ characterises the factorial part and $\tilde{\mathbf{Z}}$ the regression part.
As usual for a regression model, we assume that the errors in $\boldsymbol{\epsilon}_{ij}$ are independent with $\E(\boldsymbol{\epsilon})=\mathbf{0}$ and that a group-specific covariance matrix $\boldsymbol{\Sigma}_i:=\cov(\boldsymbol{\epsilon}_{ij}),\,i\in\{1,\dots,k\},\,j\in\{1,\dots,n_i\}$ exists.
These assumptions will later be part of Assumption (M1).
From this it follows that the observations in $\mathbf{Y}$ are assumed to be independent and identically distributed per group.
As the regression coefficients in the vector $\boldsymbol{\nu}$ depends on the dimension $\ell$, $\ell\in\{1,\dots,d\}$, but not on the group $i$, $i\in\{1,\dots,k\}$, they do not allow unequal regressions coefficients for different groups.
Allowing for unequal regression coefficients leads to uninterpretable coefficients as the magnitude of the treatment effect is not the same at different levels of $\mathbf{Z}$ \parencite{huitema_analysis_2011}.
We also refer to Figure 11.1 in \textcite{huitema_analysis_2011}. 
The covariance of $\boldsymbol{\epsilon}$ 
is given by $\boldsymbol{\Sigma}:=\cov(\boldsymbol{\epsilon})=\bigoplus_{i=1}^k(\mathbf{I}_{n_i}\otimes\mathbf{\Sigma}_i)$.
As suggested in \textcite{zimmermann_multivariate_2020}, the vector $\boldsymbol{\mu}$ can be estimated by the ordinary least squares (OLS) estimator $\hat{\boldsymbol{\mu}}=(\boldsymbol{\mu}_1',\dots,\boldsymbol{\mu}_k')'$, that is
\begin{align}
\label{equation:covariate-adjusted-mean}
\hat{\boldsymbol{\mu}}_i=\bar{\mathbf{Y}}_{i.}-\left(\bar{\mathbf{z}}_{i.}\otimes\mathbf{1}_d'\right)\hat{\boldsymbol{\nu}},
\end{align}
where the dot notation means averaging over all subjects in group $i$.
The vector $\hat{\boldsymbol{\nu}}=(\hat{\boldsymbol{\nu}}_{1}',\dots,\hat{\boldsymbol{\nu}}_{c}')'$ is the OLS estimator of $\boldsymbol{\nu}$, where $\hat{\boldsymbol{\nu}}_{m}=(\hat\nu_{m1},\dots,\hat\nu_{md})'$ for every $m\in\{1,\dots,c\}$.
To define $\hat{\boldsymbol{\nu}}$ in matrix notation we use the matrices $\mathbf{M}=\bigoplus_{i=1}^k\mathbf{1}_{n_i}$, $\mathbf{P}_M:=\mathbf{M}(\mathbf{M}'\mathbf{M})^{-1}\mathbf{M}'$ and $\mathbf{W}:=(\mathbf{I}_n-\mathbf{P}_M)\mathbf{Z}$ and define $\hat{\boldsymbol{\nu}}=[(\mathbf{W}'\mathbf{W})^{-1}\mathbf{W}'\otimes\mathbf{I}_d]\mathbf{Y}$ 
, where $(\mathbf{I}_n-\mathbf{P}_M)$ adjusts the covariates $\mathbf{Z}$ in such a way that they are correctly multiplied with the factorial part and the classical multivariate OLS estimator is calculated by $\mathbf{W}$.
To see the connection with the classical formulation of linear models, the OLS estimator for $\boldsymbol{\beta}=(\boldsymbol{\mu}',\boldsymbol{\nu}')'$ may also be written as $\hat{\boldsymbol{\beta}}=(\hat{\boldsymbol{\mu}}',\hat{\boldsymbol{\nu}}')'=(\tilde{\mathbf{X}}'\tilde{\mathbf{X}})^{-1}\tilde{\mathbf{X}}'\mathbf{Y}$.

In order to show the asymptotic behaviour of this estimator and to derive asymptotic MCTPs based thereon, the following assumptions are made \parencite{zimmermann_multivariate_2020}, where  
here and throughout all convergences are understood as $n\to\infty$.
\begin{itemize}
\item[(M1)] The errors $\boldsymbol{\epsilon}_{ij}$ are independent and identically distributed in every group $i\in\{1,$ $\dots,k\}$ with $\E(\boldsymbol{\epsilon_{ij}})=\mathbf{0}$, $\cov(\boldsymbol{\epsilon}_{ij})=\boldsymbol{\Sigma}_i$ and $\E(\|\boldsymbol{\epsilon}_{ij}\|^4) < \infty$ for all $i\in\{1,\dots,k\}$ and $j\in\{1,\dots,n_i\}$.
\item[(M2)] The variance is positive, i.e. $\sigma_{i\ell}^2:=\var(\epsilon_{ij\ell})>0$ for all $i\in\{1,\dots,k\}$ and $\ell\in\{1,\dots,d\}$.
\item[(M3)] The groups do not vanish, i.e. $n_i/n\to\kappa_i>0$.
\item[(M4)] The matrix of covariates $\mathbf{Z}$ has full column rank, i.e. the columns of $\mathbf{Z}$ are linearly independent of each other, and they should be independent of the columns of $\bigoplus_{i=1}^k\mathbf{1}_{n_i}$.
\item[(M5)] $1/n_i\sum_{j=1}^{n_i}z_{ijm}\to \pi_{im}\in\br$ for all $i\in\{1,\dots,k\}$ and $m\in\{1,\dots,c\}$.
\item[(M6)] $1/n_i\sum_{j=1}^{n_i}\mathbf{z}_{ij}\mathbf{z}_{ij}'\to\Pi_i\in\br^{c\times c}$ for all $i\in\{1,\dots,k\}$. 
\end{itemize}
Note that (M1) does not postulate a specific distribution class (such as normality). 
In particular, the distributions can differ between groups.
Singularity of the covariance matrix $\mathbf{\Sigma}$ is allowed through (M2).
Assumption (M3) is a standard assumption in asymptotic frameworks with several groups while (M4) avoids collinearity.
As stated in \textcite{zimmermann_multivariate_2020}, (M5) and (M6) have technical reasons.

We need the following central limit theorem, to derive the asymptotic behaviour of test statistics.
It is proven in the Appendix of \textcite{zimmermann_multivariate_2020}, see (A1) in the Proof of Theorem 1 therein. 
We state it here as a separate theorem.

\begin{prop}\label{thm:clt}
Let $\boldsymbol{\beta}=(\boldsymbol{\mu}',\boldsymbol{\nu}')'$ and $\hat{\boldsymbol{\beta}}=(\hat{\boldsymbol{\mu}}', \hat{\boldsymbol{\nu}}')'$. 
Then, under (M1), (M3), (M4), (M5) and (M6) it holds
\begin{align*}
\sqrt{n}\left(\hat{\boldsymbol{\beta}}-\boldsymbol{\beta}\right)\stackrel{d}{\longrightarrow}\mathbf{N}\sim \mathcal{N}(\mathbf{0}_{d\,(c+k)},\boldsymbol{\Lambda}),
\end{align*}
where $\boldsymbol {\Lambda}:=\lim_{n\to\infty}n(\tilde{\textbf{X}}'\tilde{\textbf{X}})^{-1}\tilde{\textbf{X}}'\boldsymbol{\Sigma}\tilde{\textbf{X}}(\tilde{\textbf{X}}'\tilde{\textbf{X}})^{-1}$. 
\end{prop}

\section{Statistical Methods}\label{sec:methods}
\subsection{Heteroscedasticity- and Singularity-robust Estimation of $\mathbf{\Lambda}$}\label{sec:covestimation}

For statistical inference we are not only interested in an estimator for $\boldsymbol{\beta}$, but also in an estimator for $\boldsymbol{\Lambda}$, the limiting covariance matrix of $\boldsymbol{\beta}$.
Consider the vector of residuals $\hat{\boldsymbol{\epsilon}}= (\hat{\boldsymbol{\epsilon}}_1',\dots,\hat{\boldsymbol{\epsilon}}'_k)'=\mathbf{Y}-\tilde{\mathbf{M}}\hat{\boldsymbol{\mu}}-\tilde{\mathbf{Z}}\hat{\boldsymbol{\nu}}$ and, more precisely, for subject $j$ in group $i$ the residual $\hat{\boldsymbol{\epsilon}}_{ij}$ $=\mathbf{Y}_{ij}-\hat{\boldsymbol{\mu}}_i-(\mathbf{z}_{ij}'\otimes\mathbf{I}_d)\hat{\boldsymbol{\nu}}$ $ \in\br^d$.
We then define the squared residuals as
\begin{align}\label{equ:sqres}
\hat{\boldsymbol{\Sigma}}_{ij}=\hat{\boldsymbol{\epsilon}}_{ij}\hat{\boldsymbol{\epsilon}}_{ij}'.
\end{align}
We now use the multivariate generalisation of the classical regression covariance estimator by \textcite{eicker_asymptotic_1963}, i.e. we consider an adjusted block diagonal matrix of the matrices (\ref{equ:sqres}) as the mid part of a sandwich covariance estimator. 
As our model allows for heteroscedasticity, we adjust the sandwich estimator for that. 
Classical adjustments for heteroscedasticity are adaptable for multivariate outcomes and do not influence the statistical inference of this model as they converge to $1$, see \textcite{zimmermann_multivariate_2020, welz2023cluster}. 

Therefore, all of them are applicable in this framework.
To follow the recommendation of \textcite{zimmermann_multivariate_2020}, we propose a multivariate generalization of the HC4-adjustment \parencite{cribari-neto_asymptotic_2004}, which means that $(1-p_{ij})^{-\delta_{ij}/2}$  is multiplied with every squared residual $\hat{\boldsymbol{\Sigma}}_{ij}$, where $p_{ij}$ is the $(i,j)$-th diagonal element of the matrix $\mathbf{X}(\mathbf{X}'\mathbf{X})^{-1}\mathbf{X}' $, $i\in\{1,\dots,k\},\,j\in\{1,\dots,n_i\}$ with $\delta_{ij}:=\min\{4,p_{ij}/(n^{-1}\sum_{r=1}^k\sum_{s=1}^{n_r}p_{rs}\}$.
In the general ANCOVA model studied in \textcite{zimmermann_small-sample_2019}, the HC4-adjusted covariance estimator was shown to be most effective compared to other heteroscedasticity-consistent approaches.
Therefore, we consider the adjusted squared residuals
\begin{align*}
\hat{\boldsymbol{\Sigma}}^H_{ij}=(1-p_{ij})^{-\delta_{ij}}\hat{\boldsymbol{\Sigma}}_{ij}
\end{align*}
and the block diagonal matrix $\hat{\boldsymbol{\Sigma}}=\bigoplus_{i=1}^k\bigoplus_{j=1}^{n_i}\hat{\boldsymbol{\Sigma}}^H_{ij}$ for all $i\in\{1,\dots,k\}$ and $j\in\{1,\dots,n_i\}$.
If this matrix is used as the center matrix of a sandwich covariance estimator \parencite{eicker_asymptotic_1963}, the upper left $dk\times dk$ dimensional matrix block of
\begin{align}\label{eq:sandwich}
\hat{\boldsymbol{\Lambda}}=n(\tilde{\mathbf{X}}'\tilde{\mathbf{X}})^{-1}\tilde{\mathbf{X}}'\hat{\boldsymbol{\Sigma}}\tilde{\mathbf{X}}(\tilde{\mathbf{X}}'\tilde{\mathbf{X}})^{-1}
\end{align}
is a consistent estimator for $\boldsymbol{\Lambda}_{11}$, the upper left block of $\boldsymbol{\Lambda}$, and is defined as $\hat{\boldsymbol{\Lambda}}_{11}$. 
To be robust against singularity, we follow the idea of \textcite{friedrich_mats_2018}, and only use the diagonal elements of 
 (\ref{eq:sandwich}) as covariance matrix estimator and zeros otherwise. 
We combine both ideas to get an heteroscedasty and singularity robust covariate adjusted covariance estimator by
\begin{align*}
\hat{\mathbf{D}}:=\left(\hat{\boldsymbol{\Lambda}}_{11}\right)_0,
\end{align*}
where the subscript zero means that only the diagonal elements are kept while all other matrix elements (on the off-diagonal) are set to zero.
Combining \textcite{eicker_asymptotic_1963} with the Assumptions (M1)- (M6) implies that the estimators $\hat{\boldsymbol{\mu}}$ and $\hat{\boldsymbol{\nu}}$ are consistent for $\boldsymbol{\mu}$ and $\boldsymbol{\nu}$, respectively.
With the same argument it follows that the sandwich estimator $\hat{\boldsymbol{\Lambda}}$ 
is consistent for $\boldsymbol{\Lambda}$. 
Applying the Continuous Mapping Theorem, it thus follows that the covariance estimator $\hat{\mathbf{D}}$ is also consistent for $\mathbf{D}=(\boldsymbol{\Lambda}_{11})_0$, see also Theorem 1 and 2 in \textcite{zimmermann_multivariate_2020} for similar results. 

\subsection{Multiple Testing Problem}\label{sec:mtestprob}
Within this framework we are able to formulate multiple testing problems consisting of $r$ tests regarding the vector of adjusted means $\boldsymbol{\mu}$. 
To this end we consider $r$ contrast vectors $\mathbf{h}_s=(\mathbf{h}'_{s1},\dots,\mathbf{h}'_{sk})'\in\br^{kd}$, $\mathbf{h}_{si}=(h_{si1},\dots,h_{sid})'$ for all $s\in\{1,\dots,r\}$.
The vector $\mathbf{h}_s$ is a contrast vector iff $\sum_{i=1}^k\sum_{\ell=1}^dh_{si\ell}=0$, and we combine all of them in the contrast matrix $\mathbf{H}=(\mathbf{h}_1,\dots,\mathbf{h}_r)'\in\br^{r\times kd}$.
Then, the local hypotheses are $\mathcal{H}_{0,s}:\,\mathbf{h}_s'\boldsymbol{\mu}=\mathbf{0}$ for all $s\in\{1,\dots,r\}$, defining a family of local null hypotheses 
\begin{align}\label{eq:globaltest}
\Omega=\left\{\mathcal{H}_{0,s}:\,\mathbf{h}_s'\boldsymbol{\mu}=\mathbf{0},\; s\in\{1,\dots,r\}\right\}.
\end{align}
This family corresponds to the global hypothesis $\mathcal{H}_0:\,\mathbf{H}\boldsymbol{\mu}=\mathbf{0}$. 
That $\mathbf{h}_s$ has to be a contrast vector is not necessary for mathematical reasons, but contrast vectors characterise the questions of interest.

\paragraph{Examples for Covered Multiple Testing Problems.}
The chosen family of hypotheses $\Omega$ covers various multiple testing problems of interest.
In fact, we are able to consider the multivariate and covariate adjusted versions of well-known multiple testing problems \parencite[cf.][]{bretz_multiple_2011,konietschke_are_2013} by choosing different matrices $\mathbf{H}$.
To build a hypothesis matrix for a multivariate testing problem from the respective univariate one $\mathbf{H}_u$ we simply use the Kronecker product of $\mathbf{H}_u$ and $\mathbf{I}_d$, that is $\mathbf{H}=\mathbf{H}_u\otimes\mathbf{I}_d$.
For example, there can be realized the following multivariate testing procedures with this technique:
\begin{enumerate}
\item \textbf{Multiple testing in a multivariate two-sample problem.}
For $k=2$ groups and an arbitrary dimension $d$ we can compare the adjusted means $\mu_{i,\ell}$ for every endpoint: $\mathcal{H}_{0,\ell}:\,\mu_{1\ell}=\mu_{2,\ell}$, $\ell\in\{1,\dots,d\}$.
This multiple testing problem occurs if multiple correlated endpoints have to be analysed and the dimension-wise hypotheses are of interest.
The hypothesis can be defined by the matrix $\mathbf{H}=(1,-1)\otimes\mathbf{I}_d$.
This testing problem is considered  in the data example in Section \ref{sec:dataanalysis}.
\item \textbf{Multivariate many-to-one comparison.}
For arbitrary $k$ and $d$ the Kronecker product of the Dunnett-type matrix \parencite{dunnett_multiple_1955} and $\mathbf{I}_d$ leads to the hypotheses $\mathcal{H}_{0,i\ell}:\,\mu_{1\ell}=\mu_{i\ell},\,i\in\{2,\dots,k\},\,\ell\in\{1,\dots,d\}$, which compares component-wise the adjusted means of group $1$ with the adjusted mean of all other groups.
\item \textbf{Multivariate all-pair comparison.}
The equivalent use of the Tukey-type matrix \parencite{tukey_problem_1994} gives the family of hypotheses including all pair-wise comparisons: $\mathcal{H}_{0,i_1i_2\ell}:\,\mu_{i_1\ell}=\mu_{i_2\ell}, i_1,i_2\in\{1,\dots,k\},\,i_1\neq i_2,\,\ell\in\{1,\dots,d\}$.
\item \textbf{Multivariate grand-mean comparison.}
The choose of the Grand-mean-type matrix introduced by \textcite{djira_detecting_2009} leads to a component-wise comparison of the adjusted group means with the overall mean of the group-wise adjusted means $\bar\mu_{\ell}:=k^{-1}\sum_{i=1}^k\mu_{i\ell}$: $\mathcal{H}_{0,i}:\,\mu_{i\ell}=\bar{\mu}_{\ell}, i\in\{1,\dots,k\},\,\ell\in\{1,\dots,d\}$.
\end{enumerate}

To infer the local null hypothesis $\mathcal{H}_{0,s}$, we consider the local test statistics
\begin{align}\label{equ:teststatistic}
A_n(\mathbf{h}_s)=\sqrt{n}\frac{\mathbf{h}'_s\hat{\boldsymbol{\mu}}}{\sqrt{\mathbf{h}'_s\hat{\mathbf{D}}\mathbf{h}_s}}.
\end{align}
For simultaneously testing (\ref{eq:globaltest}) while adjusting for multiplicity, we must analyse the joint distribution of these statistics, i.e. 
the distribution of the vector $(A_n(\mathbf{h}_1),\dots,A_n(\mathbf{h}_r))'=
(\mathbf{H}\hat{\mathbf{D}}\mathbf{H})_0^{-1/2}\sqrt{n}\mathbf{H}\hat{\boldsymbol{\mu}} = \mathbf{A}_n(\mathbf{H})$.
Its asymptotic distribution is given in the following theorem:
\begin{satz}\label{thm:testdistribution}
Under the Assumption of the semiparametric MANCOVA model (M1)-(M6), it holds:
\begin{enumerate}
\item  Under $\mathcal{H}_0:\,\mathbf{H}\boldsymbol{\mu}=\mathbf{0}$ the vector of test statistics $\mathbf{A}_n(\mathbf{H})$ converges in distribution to a multivariate normal distribution, i.e.
\begin{align*}
\mathbf{A}_n(\mathbf{H})=(A_n(\mathbf{h}_1),\dots,A_n(\mathbf{h}_r))'\stackrel{d}{\longrightarrow}\mathbf{B},
\end{align*}
where $\mathbf{B}=(B_1,\dots,B_r)$ is $r$-dimensional and  has the expectation $\E(\mathbf{B})=\mathbf{0}$ and the covariance matrix 
\begin{align}\label{equ:multtestdistr}
\mathbf{R}:=\cov\left(\mathbf{B}\right)=\left(\mathbf{H}\mathbf{D}\mathbf{H}'\right)_0^{-\frac{1}{2}}\mathbf{H}\boldsymbol{\Lambda}_{11}\mathbf{H}'\left(\mathbf{H}\mathbf{D}\mathbf{H}'\right)_0^{-\frac{1}{2}}.
\end{align}
\item Under $\mathcal{H}_1:\,\mathbf{H}\boldsymbol{\mu}\not=\mathbf{0}$, $\mathbf{A}_n(\mathbf{H})$ converges in probability to $\infty$.
\end{enumerate} 
\end{satz}

The proof of Theorem \ref{thm:testdistribution} and all other proofs can be found in the Appendix \ref{sec:proofs}.
It is a consequence of Theorem \ref{thm:testdistribution} that $B_s,\,s\in\{1,\dots,r\}$, are normally distributed: 
\begin{align}\label{equ:testdistr}
B_s\sim\mathcal{N}\left(0,\frac{\mathbf{h}_s'\boldsymbol{\Lambda}_{11}\mathbf{h}_s}{\mathbf{h}_s'\mathbf{D}\mathbf{h}_s}\right).
\end{align}
If we would follow the traditional approach of constructing MCTPs, we would use the asymptotic distribution of $\mathbf{B}$ to derive multivariate equicoordinate quantiles, and to define local and global test decisions.
This idea follows from the union-intersection principle \parencite{roy_heuristic_1953} and was made numerically available by \textcite{bretz_numerical_2001}.
Due to standardization of the individual test statistics, a multivariate equicoordinate quantile, with equal values $q_\gamma$ in all dimensions can be defined. 
Thereby, $q_\gamma$ is the $\gamma$-quantile of the maximum tests statistic $\max_{s\in\{1,\dots,r\}}|A_n(\mathbf{h}_s)|$ at the same time.
However, this technique is not feasible in our framework since the test statistics $A_n(\mathbf{h}_s)$ have limiting distributions depending on $\mathbf{h}_s$, $s\in\{1,\dots,r\}$.
Consequently, it is not possible to compute equicoordinate quantiles. 
To overcome this problem we consider the idea of \textcite{munko_rmst-based_2024} to adjust the level of significance for local tests such that the global level is controlled and the different distributions are taken into account. 
It is based on the concept of simultaneous confidence bands of \textcite{buhlmann_sieve_1998}. 
For the adaption of this approach, we additionally consider asymptotically correct bootstrap methods presented in the following section.

We note, that it is possible to consider asymptotically valid MCTPs with equicoordinate quantiles for the covariate-adjusted mean by using $\hat{\boldsymbol{\Lambda}}_{11}$ instead of $\hat{\boldsymbol{D}}$. 
In the Supplementary Material we explain this in detail, and also discuss that this can be seen as an extension of the MCTPs from \textcite{hasler_multiple_2014} to multiple endpoints. 
However, these asymptotic MCTPs have the disadvantage, that they additionally require positive definite covariances $\boldsymbol{\Lambda}_i$, $i\in\{1,\dots,k\}$, which is a stronger assumption than (M2). 
As described in \textcite{friedrich_mats_2018} and \textcite{zimmermann_multivariate_2020} singularity can occur in multivariate data if the outcome vector has strong linear dependencies. 
That is why we opted to focus on this more general framework.

\subsection{Bootstrapping}\label{sec:boot}
To obtain suitable resampling methods we adapt the two bootstrap methods of \textcite{zimmermann_multivariate_2020}, wild and a parametric bootstrap.
The idea of the wild bootstrap is to produce variation by multiplying random variables to the residuals.
This is why we first have to generate $n$ independent identically distributed random variables $T_{ij}$ independently from the data with $\E(T_{11})=0,\,\var(T_{11})=1$ and $\sup_{i,j}\E(T^4_{ij})<\infty$ for $i\in\{1,\dots,k\}$ and $j\in\{1,\dots,n_i\}$.
Methodologically, we can consider some $T_{ij}$ which fulfils this conditions, but practically we have to decide for some specific ones.
As they turned out to be successful in \textcite{zimmermann_multivariate_2020} and the choice of weights do not seem to have much influence on the methods performance, we choose Rademacher random variables for $T_{ij}$, which means that $\prob(T_{11}=-1)=\prob(T_{11}=1)=1/2$.
Then, the elements of the wild bootstrap sample are defined as
\begin{align*}
\mathbf{Y}_{ij}^*:=\frac{T_{ij}}{\sqrt{1-p_{ij,ij}}}\hat{\boldsymbol{\epsilon}}_{ij}
\end{align*}
for every $i\in\{1,\dots,k\}$ and $j\in\{1,\dots,n_i\}$.
This process is carried out for every subject and does not depends on the component $\ell\in\{1,\dots,d\}$, which receives the dependence structure within the subjects.
With the bootstrap sample we can generate the wild bootstrap OLS estimator $\hat{\boldsymbol{\beta}}^*:=(\hat{\boldsymbol{\mu}}^{*\prime},\hat{\boldsymbol{\nu}}^{*\prime})'=(\tilde{\mathbf{X}}'\tilde{\mathbf{X}})^{-1}\tilde{\mathbf{X}}'\mathbf{Y}^*$ and the wild bootstrap covariance estimator 
\begin{align*}
\hat{\mathbf{D}}^*:=\left(\hat{\boldsymbol{\Lambda}}^*_{11}\right)_0=\left(n(\tilde{\mathbf{X}}'\tilde{\mathbf{X}})^{-1}\tilde{\mathbf{X}}'\hat{\boldsymbol{\Sigma}}^*\tilde{\mathbf{X}}(\tilde{\mathbf{X}}'\tilde{\mathbf{X}})^{-1} \right)_0,
\end{align*}
where $\hat{\boldsymbol{\Sigma}}^*=\bigoplus_{i=1}^k\bigoplus_{j=1}^{n_i}\hat{\boldsymbol{\Sigma}}^{*H}_{ij}$ and $\hat{\boldsymbol{\Sigma}}^{*H}_{ij}$ are the heteroscedasticity-robust squared residuals calculated with the wild bootstrap data for every $i\in\{1,\dots,k\}$ and $j\in\{1,\dots,n_i\}$. 
From the second part of Theorem 2 in \textcite{zimmermann_small-sample_2019} the consistency of $\hat{\mathbf{D}}^*$, i.e., $\hat{\mathbf{D}}^*\stackrel{P}{\to}\mathbf{D}$  can be concluded, because they argue that the wild bootstrap residuals $\hat{\boldsymbol{\epsilon}}^*_{ij}$ are consistent.
To define a wild bootstrap version of the test statistic $A^*_n(\mathbf{h}_s)$ we simply use the wild bootstrap estimators $\boldsymbol{\mu}^*$ and $\hat{\mathbf{D}}^*$ instead of the original estimators in formula (\ref{equ:teststatistic}), $s\in\{1,\dots,r\}$:
\begin{align}\label{equ:wildteststatistic}
A^*_n(\mathbf{h}_s)=\sqrt{n}\frac{\mathbf{h}_s'\hat{\boldsymbol{\mu}}^*}{\sqrt{\mathbf{h}_s'\hat{\mathbf{D}}^*\mathbf{h}_s}}.
\end{align}
Note that the $s$ test statistics are calculated with the same bootstrap sample.
The following Theorem states that the asymptotic distribution of the wild bootstrap test statistic $A^*_n(\mathbf{h}_s)$ is the same as the distribution of $A_n(\mathbf{h}_s)$ under $\mathcal{H}_{0,s}$.
\begin{satz}\label{thm:wildteststatistic}
Let $s\in\{1,\dots,r\}$.
The wild bootstrap test statistic $A^{\star}_n(\mathbf{h}_s)$ given in (\ref{equ:wildteststatistic}) converges conditionally given the data in distribution to the real-valued random vector $\mathbf{B}$, which characterises also the asymptotic distribution of $A_n(\mathbf{h}_s)$ under $\mathcal{H}_{0,s}$ (see Theorem \ref{thm:testdistribution}), i.e.,
\begin{align}\label{equ:convergence_wild}
\sup\left\vert \prob\left(A^*_n\left(\mathbf{h}_s\right)\le x|\mathbf{Y}\right)-\prob\left(\mathbf{B}\le x\right)\right\vert\stackrel{p}{\longrightarrow}0.
\end{align}
\end{satz}

The idea of the parametric bootstrap approach is to estimate the group-wise covariance and use these estimators to generate independent $d$-dimensional observation vectors from the normal distribution:
\begin{align*}
\mathbf{Y}_{ij}^{\star}\sim\mathcal{N}\left(\mathbf{0},\hat{\boldsymbol{\Sigma}}_i\right),
\end{align*}
for every group $i\in\{1,\dots,k\}$ and every subject $j\in\{1,\dots,n_i\}$, where
\begin{align*}
\hat{\boldsymbol{\Sigma}}_i=\frac{1}{n_i-c-1}\sum_{j=1}^{n_i}\hat{\boldsymbol{\epsilon}}_{ij}\hat{\boldsymbol{\epsilon}}_{ij}'
\end{align*}
for every group $i\in\{1,\dots,k\}$.
Analogously to the wild bootstrap approach we can calculate the parametric bootstrap versions $\hat{\boldsymbol{\beta}}^{\star}$ and $\hat{\mathbf{D}}^{\star}$ of the estimators and the test statistic
\begin{align}\label{equ:parateststatistic}
A^{\star}_n(\mathbf{h}_s)=\sqrt{n}\frac{\mathbf{h}_s'\hat{\boldsymbol{\mu}}^{\star}}{\sqrt{\mathbf{h}_s'\hat{\mathbf{D}}^{\star}\mathbf{h}_s}}.
\end{align}
for $s\in\{1,\dots,r\}$.
Again, the $s$ test statistics are calculated from the same bootstrap sample.
In the Proof of Theorem 4 of \textcite{zimmermann_multivariate_2020}, it is shown that $\hat{\mathbf{D}}^{\star}$ is a consistent estimator for $\mathbf{D}$.
And similar to the wild bootstrap approach we are able to state that the asymptotic distribution of the parametric bootstrap test statistic $A^{\star}_n(\mathbf{h}_s)$ is the same as the distribution of $A_n(\mathbf{h}_s)$ under $\mathcal{H}_{0,s}$.
\begin{satz}\label{thm:parateststatistic}
Let $s\in\{1,\dots,r\}$.
The parametric bootstrap test statistic $A^{\star}_n(\mathbf{h}_s)$ given in (\ref{equ:parateststatistic}) converges conditionally given the data in distribution to the real-valued random vector $\mathbf{B}$, which characterises also the asymptotic distribution of $A_n(\mathbf{h}_s)$ under $\mathcal{H}_{0,s}$  (see Theorem \ref{thm:testdistribution}), i.e.,
\begin{align}\label{equ:convergence_para}
\sup\left\vert \prob\left(A^{\star}_n\left(\mathbf{h}_s\right)\le x|\mathbf{Y}\right)-\prob\left(\mathbf{B}\le x\right)\right\vert\stackrel{p}{\longrightarrow}0.
\end{align}
\end{satz}

\subsection{Determination of Critical Values}\label{sec:critvalues}
To get suitable critical values for the test decision in the semiparametric MANOVA model we draw $B$ bootstrap samples with one of the methods above.
Consider for every sample $b\in\{1,\dots,B\}$ the vector of test statistics $(A^{\circ,b}_n(\mathbf{h}_1),\dots,A^{\circ,b}_n(\mathbf{h}_r))$, $\circ\in\{*,\star\}$.
Define $q^{\circ}_{s,1-\gamma}$, the $(1-\gamma)$-quantile of $\vert A^{\circ,1}_n(\mathbf{h}_s)\vert,\dots,\vert A^{\circ,B}_n(\mathbf{h}_s)\vert, \,s\in\{1,\dots,r\},\,\circ\in\{*,\star\}$.
The idea of \textcite{munko_rmst-based_2024} is to adjust the significance level $\gamma$ for each local test such that the level $\alpha$ is controlled globally.
For that, define the estimated family-wise type I error rate with the critical value $q^{\circ}_{r,\gamma}$:
\begin{align*}
\text{FWER}^{\circ}_n(\gamma):=\frac{1}{B}\sum_{b=1}^B\mathds{1}\left\{\exists\;s\in\{1,\dots,r\}:\vert A^{\circ,b}_n(\mathbf{h}_s)\vert>q^{\circ}_{s,1-\gamma}\right\}
\end{align*}
for $\circ\in\{*,\star\}$ and $\gamma\in[0,1]$.
Then, \textcite{munko_rmst-based_2024} define the adjusted level $\gamma_n(\alpha)$ as:
\begin{align}
	\label{eq:adjusted-level-bootstrap-mctp}
	\gamma_n(\alpha):=\max\left\{\gamma\in\left\{0,\frac{1}{B},\dots,\frac{B-1}{B}\right\}|\text{FWER}^{\circ}_n(\gamma)\le\alpha\right\},
\end{align}
which means that $\gamma_n(\alpha)$ is the largest value such that $\text{FWER}^{\circ}_n(\gamma)$ is bounded by the global level of significance $\alpha$.
Here, the maximum is evaluated over the set $\{0,1/B,\dots,(B-1)/B\}$, because the quantiles can only take $B$ different values.
With the adjusted level $\gamma_n(\alpha)$ we are able to formulate some decision rules regarding the local and global hypotheses for the two different bootstrap approaches.
For every $s\in\{1,\dots,r\}$ and $\circ\in\{*,\star\}$ the hypothesis $\mathcal{H}_{0,s}$ of $\Omega$ is rejected if and only if $|A_n(\mathbf{h}_s)|>q^{\circ}_{s,1-\gamma_n(\alpha)}$ or equivalently $|A_n(\mathbf{h}_s)|/q^{\circ}_{s,1-\gamma_n(\alpha)}>1$, i.e. we can define tests
\begin{align*}
\varphi^{\circ}_{n,s}=\mathds{1}\{|A_n(\mathbf{h}_s)|>q^{\circ}_{s,1-\gamma_n(\alpha)}\}.
\end{align*}
It is possible to construct simultaneous confidence intervals from the multiple testing procedure for $\mathbf{h}'_s\boldsymbol{\mu}$ with the global confidence level $1-\alpha$:
\begin{align*}
\left[\mathbf{h}'_s\hat{\boldsymbol{\mu}}\pm q^{\circ}_{s,1-\gamma_n(\alpha)}\frac{\sqrt{\mathbf{h}'_s\hat{\mathbf{D}}\mathbf{h}_s}}{\sqrt{n}}\right].
\end{align*}
In line with the classic MCTPs the global hypothesis $\mathcal{H}_0$ is rejected, if and only if at least one $\mathcal{H}_{0,s}$ is rejected.
This leads to the global test
\begin{align*}
\varphi^{\circ}_{n}=\max_{s\in\{1,\dots,r\}}\mathds{1}\left\{\frac{|A_n(\mathbf{h}_s)|}{q^{\circ}_{s,1-\gamma_n(\alpha)}}>1\right\},\,\circ\in\{*,\star\}.
\end{align*}
This formulation incorporates the different distributions of $A_n(\mathbf{h}_s),\,s\in\{1,\dots,r\}$ by considering individual quantiles $q^{\circ}_{s,1-\gamma_n(\alpha)}$ as critical values, but uses the same level of significance $\gamma_n(\alpha)$ for all tests.
To ensure, that the level of significance of the global test and the family-wise type I error rate of $\Omega$ is controlled asymptotically we state the following Theorem:
\begin{satz}\label{thm:adjlevel}
Let $T\subset\{1,\dots,r\}$ denote the subset of true hypotheses $\mathcal{H}_{0,s}$ of $\Omega$.
Then, $\varphi^{\circ}_n,\,\circ\in\{*,\star\}$ is an asymptotic level-$\alpha$ test, i.e. with $B=B(n)\to\infty$ as $n\to\infty$, it holds that
\begin{align*}
\lim_{n\to\infty}\prob\left(\exists\,s\in T:\,|A_n(\mathbf{h}_s)|>q^{\circ}_{s,1-\gamma_n(\alpha)}\right)\le\alpha,
\end{align*}
where equality holds, if $T=\{1,\dots,r\}$.
\end{satz}
Theorem \ref{thm:wildteststatistic} and Theorem \ref{thm:parateststatistic} state that the conditional distributions of $A^{\circ}_n(\mathbf{h}_s),\,\circ\in\{*,\star\}$ are asymptotically the same as the distribution of $A_n(\mathbf{h}_s)$ under $\mathcal{H}_{0,s}$.
Under $\mathcal{H}_{1,s}$ we have the situation that the test statistic $A_n(\mathbf{h}_s)$ is divergent and and $A^{\circ}_n(\mathbf{h}_s),\,\circ\in\{*,\star\}$ has still the same normal distribution.
Therefore, the tests $\varphi^{\circ}_{n,s}$ and $\varphi^{\circ}_n$ are consistent.

We considered also a version of this tests where we do not assume that the test statistics $A^{\circ}_n(\mathbf{h}_s),\,\circ\in\{*,\star\},$ are symmetrically distributed and therefore do not use the absolute value of the test statistics, but two asymmetric critical values from an optimized local level that takes the asymmetry into account.
It turns out that this method is not superior to the symmetric version.
That is why we state the methodology and some simulation results in the Supplementary Material.

\subsection{P-Values}
\label{sec:p-values}

In this section, we are going to introduce local and global p-values for the bootstrap MCTPs. 
For the asymptotic MCTPs sketched in \autoref{sec:mtestprob}, we refer to the Supplementary Material.
Moreover, we are going to show that the comparison of these p-values against an adjusted significance level provides an equivalent way of expressing both the local test $\varphi^{\circ}_{n, s}$ and the global test $\varphi^{\circ}_{n}$.
Comparing p-values with an adjusted level of significance comes with a noteworthy advantage: while the test decision of $\varphi^{\circ}_{n, s}$ through the critical values is based on a different value $q^{\circ}_{s,1-\gamma_n(\alpha)}$ for each test ${s \in \{1, \ldots r\}}$, p-value-based test decisions require only one adjusted significance level for all tests.
In case of the bootstrap MCTPs, this adjusted significance level is defined as $\gamma_n(\alpha)$.
In contrast to classical p-values, it has to be calculated dependently from $\alpha$.
Notably, the consideration of an adjusted level of significance is conceptually analogous to Bonferroni-adjustment \parencite{dunn_multiple_1961}, where an adjusted significance level is obtained by dividing the global significance level $\alpha$ by the number of tests.
This local level can be used to get a test decision, either it is used to calculate a critical value or a p-value.
This analogy eases comparability between MCTPs and Bonferroni, and therefore, will subsequentially help us illustrating MCTPs (cf.\ \autoref{sec:dataanalysis}).

We use the definition of \textcite{munko_rmst-based_2024} to introduce local p-values for bootstrap MCTPs as follows:
\begin{align}
  \label{eq:local-p-value-bootstrap-mctp}
  p_{n, s} := \frac{1}{B} \sum_{b = 1}^{B} \mathds{1} \left\{ | A^{\circ,b}_n(\mathbf{h}_s) | \geq | A_n(\mathbf{h}_s) | \right\}
\end{align}
To give an intuition, $p_{n, s}$ is the fraction of bootstrap runs, for which the bootstrap test statistic $A^{\circ,b}_n(\mathbf{h}_s)$ is at least as large as $A_n(\mathbf{h}_s)$ in absolute values.
Recall from Theorem \ref{thm:wildteststatistic} and Theorem \ref{thm:parateststatistic} that $A^{\circ,b}_n(\mathbf{h}_s)$ asymptotically mimics the distribution of $A_n(\mathbf{h}_s)$ under $\mathcal{H}_{0,s}$.
Therefore, it is easy to see that the definition of $p_{n, s}$ is in line with the well-established interpretation of p-values as the probability that the test statistic assumes a value at least as extreme as the one observed when the null hypothesis holds \parencite[e.g.][]{woodwardEpidemiologyStudyDesign2013}.
Eventually, a global p-value is obtained by $p_n := \min \{ p_{n, 1}, \ldots, p_{n, r} \}$.
The following proposition shows equivalence to the tests defined above:
\begin{prop}\label{thm:p-value-equivalence}
\leavevmode
\begin{itemize}
  \item[(i)] For each ${s \in \{1, \ldots r\}}$, it holds  $p_{n, s} \leq \gamma_n(\alpha)$ if and only if  $\varphi^{\circ}_{n, s} = 1$,
  \item[(ii)] It holds that $p_n \leq \gamma_n(\alpha)$ if and only if $\varphi^{\circ}_{n} = 1$.
\end{itemize}
\end{prop}

\section{Simulations}
\label{sec:simus}

In order to analyse the small sample performance of the developed testing procedures we did an extensive simulation study on the Linux HPC cluster of TU Dortmund University (LiDo3) via the computing environment \textsf{R}, version 4.2.1 \textcite{r_core_team_r_2022}.
For each simulation scenario we considered $5000$ simulation runs, $2000$ bootstrap iterations and the level of significance $\alpha=0.05$.
We chose a simulation set-up similarly to \textcite{zimmermann_multivariate_2020}, accordingly, the basis of our simulations is the following data-generation process:
\begin{align*}
\mathbf{Y}_i=\boldsymbol{\mu}_i+\mathbf{Z}_i\boldsymbol{\nu}+\boldsymbol{\epsilon}_i,\quad\boldsymbol{\epsilon}_i=\boldsymbol{\Sigma}_i^{\frac{1}{2}}\mathbf{X}_i,\,i\in\{1,\dots,k\}.
\end{align*}
Here, $\mathbf{X}_i\in\br^{n_i\times d}$ is a matrix of standardized random data from different distributions: standard normal distribution (\textsf{N}), t-distribution with $3$ degrees of freedom ($t_3$), $\chi^2$-~distribution with $3$ degrees of freedom ($\chi_3^2$), logarithmic standard normal distribution (\textsf{LN}) and double exponential distribution (\textsf{DExp}).
We chose the balanced and unbalanced sample sizes $\mathbf{n}^{(1)}=(10,\dots,10)'\in\br^{k}$, $\mathbf{n}^{(2)}=(20,10,\dots,10)'\in\br^{k}$, $\mathbf{n}^{(3)}=(10,\dots,10,20)'\in\br^{k}$ and its multiples $K\cdot\mathbf{n}^{(t)},t\in\{1,2,3\}$.
The matrix $\boldsymbol{\Sigma}_i^{\frac{1}{2}}\in\br^{d\times d}$ is the square-root of a certain covariance matrix.
We consider different scenarios:
\begin{enumerate}
\item Homoscedastic covariance: $\boldsymbol{\Sigma}_i=\mathbf{I}_d+0.5\cdot(\mathbf{J}_d-\mathbf{I}_d),\,i\in\{1,\dots,k\}$.
\item Heteroscedastic covariance: $\boldsymbol{\Sigma}_i=\mathbf{I}_d+0.5\cdot(\mathbf{J}_d-\mathbf{I}_d)$ for $i\in\{1,\dots,k-1\}$ and $\boldsymbol{\Sigma}_k=2\cdot\mathbf{I}_d+0.5\cdot(\mathbf{J}_d-\mathbf{I}_d)$.
\item Singular covariance: $\boldsymbol{\Sigma}_i=\begin{cases}
\begin{pmatrix}
1 & 0.5\\
0.5 & 0.25
\end{pmatrix},\quad d=2\\
\begin{pmatrix}
6 & 3 & 3\\
3 & 2 & 3\\
3 & 3 & 6\\
\end{pmatrix},\quad d=3\\
\begin{pmatrix}
6 & 3 & 3 & 3\\
3 & 6 & 3 & 3 \\
3 & 3 & 2.5 & 3 \\
3 & 3 & 3 & 6 \\
\end{pmatrix},\quad d=4
\end{cases}$,
$i\in\{1,\dots,k\}$.
\end{enumerate}
If the second covariance setting is combined with the sample sizes $K\cdot\mathbf{n}^{(2)}$ and $K\cdot\mathbf{n}^{(3)}$ the situations of negative respective positive pairing occur, where the group with the smaller sample size has the bigger respective the smaller variance.
The number of covariates is set to $c=2$ for all simulations and the matrix of covariates $\mathbf{Z}_i=(\mathbf{z}_{i1},\mathbf{z}_{i2})\in\br^{n\times c}$ is drawn from uniform distributions separately: $\mathbf{z}_{i1}\sim\mathcal{U}((-10,10))$ and for $\mathbf{z}_{i2}$ the first half is drawn from $\mathcal{U}((0,5))$ and the second half from $\mathcal{U}((-2,-1))$.
Additionally, we accept the generated covariates only under certain conditions of dispersion, see the R-script \texttt{covariates.R} in the Supplementary Material for further details.
The regression coefficients $\boldsymbol{\nu}\in\br^{c\times d}$ are set to $\boldsymbol{\nu}=
(-0.5, \mathbf{1}'_{d-2}, -1|
1.5, 2\cdot\mathbf{1}'_{d-2}, 3)$.
This choice is in line with the simulation set-up in \textcite{zimmermann_multivariate_2020}.
For $d=2$ they did a small simulation to ensure the association between the covariates and the selected outcomes with that choice of $\boldsymbol{\nu}$.
Additionally, we did a similar simulation for dimensions $d\in\{3,4\}$ by fitting univariate models and checking if the covariates are significant at the level of $5\%$.
We chose a sample size of $40$ per group and did $1000$ simulation runs.
Especially, this choice of $\boldsymbol{\nu}$ ensures the linear relationship between the components $1$ and $d$ of the outcomes in the singular covariance setting.

In all simulation runs, we compare the method of multiple contrast test procedures with wild (\textit{MCTP-wild}) and parametric (\textit{MCTP-param}) bootstrap with the asymptotic multiple contrast test procedure explained in Section \ref{sec:mtestprob} (\textit{MCTP-norm}, \textit{MCTP-t}) and with the MANCATS by \textcite{zimmermann_multivariate_2020}.
The latter is also considered with wild (\textit{MANCATS-wild}) and parametric (\textit{MANCATS-param}) bootstrap and we made it comparable in our multiple testing problem by simply using the Bonferroni-adjustment \parencite{dunn_multiple_1961}.
Note that the asymptotic MCTPs are not defined in the singular covariance setting (3).
Therefore, we compare six methods in this simulation study.
We simulated also the asymmetric MCTPs explained in Section \ref{sec:critvalues} in all scenarios.
These methods are not considered throughout this paper but there are some analyses in the Supplementary Material.
All simulation results can be found in the Supplementary Material.

\subsection{Simulation Results Under the Null Hypothesis}
\begin{figure}
\centering
\includegraphics[scale=0.7]{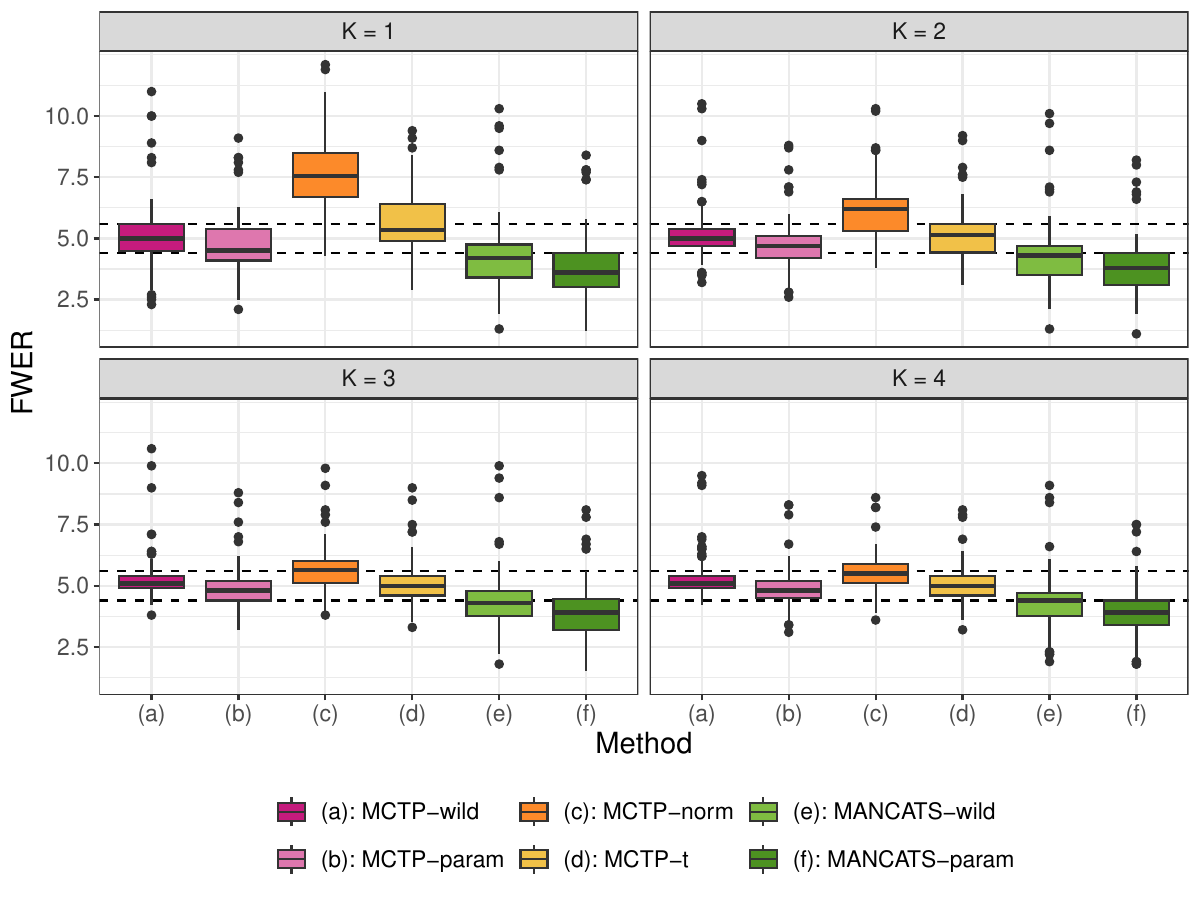}
\caption{Empirical family-wise type error rate in $\%$ for Dunnett's testing problem with $k=3$ and $d\in\{2,3,4\}$ split by the sample size multiplier $K\in\{1,2,3,4\}$, i.e., $\mathbf{n}=K\cdot\mathbf{n}^{(t)},\,t\in\{1,2,3\}$.}
\label{fig:5.2fwer_k}
\end{figure}

\begin{figure}
\centering
\includegraphics[scale=0.71]{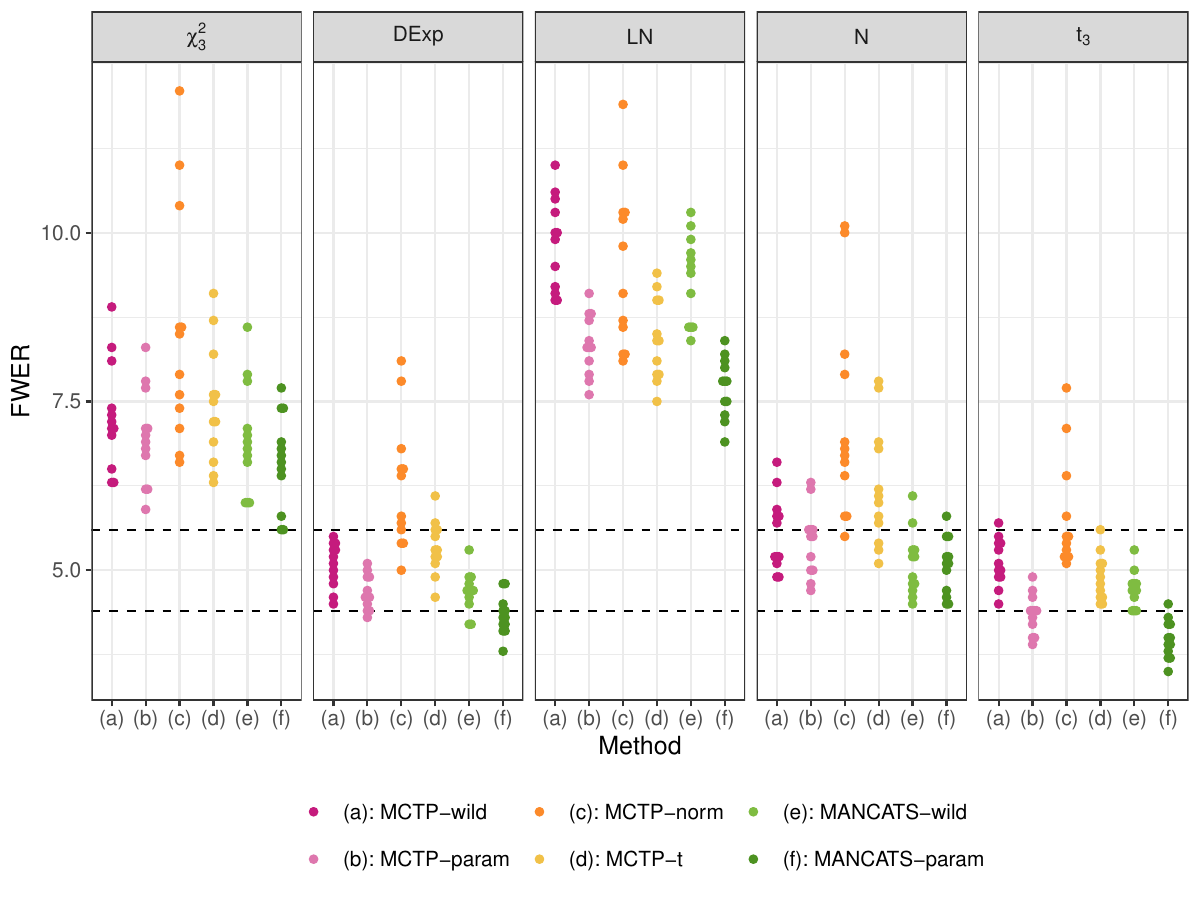}
\caption{Empirical family-wise type I error rate in $\%$ for Dunnett's testing problem with $k=3$ and $d\in\{2,3,4\}$  for the settings with negative pairing (the bigger covariance occurs in the smaller sample size for a heteroscedastic unbalanced setting) split by the distributions: standard normal distribution (\textsf{N}), t-distribution with $3$ degrees of freedom ($t_3$), $\chi^2$-distribution with $3$ degrees of freedom ($\chi_3^2$), logarithmic standard normal distribution (\textsf{LN}) and double exponential distribution (\textsf{DExp}).}
\label{fig:5.2fwer_negativepairing}
\end{figure}

For simulations under the null hypothesis we ensure that our data fulfils the global null hypothesis $\mathcal{H}_0$ and set $\boldsymbol{\mu}_i=\mathbf{0}$ for all groups $i\in\{1,\dots,k\}$.
In the analysis of the control of the family-wise type I error rate (FWER) we consider small and moderate sample sizes, consequently, the choice of the sample size multiplier $K$ is widely and the sample are multiplied with $K\in\{1,2,3,4\}$ as explained above.
To get a comprehensive overview about the FWER-control of the testing procedures, we considered various numbers of groups and dimensions as well as different testing problems.
In Table \ref{tab:simsettings}, we present the combinations we included in our simulation study.
\begin{table}
\centering
\begin{tabular}{lrr}
\toprule
Testing Problem & Groups & Dimensions\\
\midrule
Two-sample & $2$ & $2,3,4$ \\
Dunnett & $3$ & $2,3,4$ \\
Dunnett & $4$ & $2,3$ \\
Tukey & $3$ & $2,3$ \\
Tukey & $4$ & $2$ \\
\bottomrule
\end{tabular}
\caption{Combinations of considered testing problems, groups $k$ and dimensions $d$ in the simulation study regarding FWER-control. For the testing problems, see the explanation in Section \ref{sec:mtestprob}.}
\label{tab:simsettings}
\end{table}
As an example we present in the paper plots regarding the simulations settings considering Dunnett's testing problem and $d\in\{2,3,4\}$ dimensions and $k=3$ groups.
In Figure \ref{fig:5.2fwer_k} this simulations results are presented as boxplots \parencite[from \textsf{R}-package \texttt{ggplot2}, see][]{wickham_ggplot2_2016} split by the sample size multiplier $K$.
Here, it is observable that the empirical FWER of the resampling MCTPs is in the $95\%$ binomial interval $[4.4,5.6]$ for most simulation settings even for the smaller samples.
The MCTP with parametric bootstrap tends to be a little bit more conservative than the alternative with wild bootstrap in most settings.
The asymptotic MCTPs tend to be liberal for smaller sample sizes, this effect is stronger for the asymptotic MCTPs with multivariate normal distribution.
The Bonferroni-adjusted MANCATS tend to a conservative behaviour as expected.
For all methods, there are a few settings with a highly liberal behaviour.
In Figure \ref{fig:5.2fwer_negativepairing} the subgroup of settings with negative pairing are plotted as bee swarms \parencite{eklund_beeswarm_2021} split by the distributions and we can see that liberal behaviour occurs in settings with negative pairing and the $\chi_3^2$- or the logarithmic normal distribution.
A moderately liberal behaviour can also be observed in the plot for the normal distribution.
Here, the resampling MCTPs are less liberal than the asymptotic one with $t$-distribution.
This liberal behaviour is not surprising as it is also observed in other multivariate semiparametric models for factorial designs that negative pairing can be a problem, see for example \textcite{konietschke_parametric_2015}.
In the logarithmic normal and in the $t_3$-distributed settings of the presented data in Figure \ref{fig:5.2fwer_k}, the resampling MCTPs tend to a conservative behaviour (empirical FWER smaller than $4.4$).
As the MCTP with parametric bootstrap tends to be a bit more conservative than the MCTP with wild bootstrap the MCTP with parametric bootstrap produced more settings with an FWER lower than $4.4$.
In the Supplemental Material, we present all simulation results split by the distribution and the covariance settings as bee swarm plots \parencite{eklund_beeswarm_2021}.
From this plots, it can be observed that the FWER-control of all methods is not really stable for the logarithmic normal distribution with the heteroscedastic covariance.
This phenomenon is more pronounced if the dimension $d$ is higher.
It has to be pointed out that the asymptotic MCTP with multivariate $t$-distribution performs similarly well as the resampling MCTPs especially for the bigger sample sizes.
But a problem of this method is that it is not defined for singular covariance scenarios and can not be applied in these settings.
It turns out that the simulations results are not much different for other testing problems or numbers of groups and dimensions.
Nevertheless, further plots regarding simulation results can be found in the Supplementary Material.

All in all, the simulation study regarding FWER leads to a recommendation of the two resampling MCTPs.
These methods allow potential singularity and work also for small samples and in heterogeneous data, which is an advantage compared to the asymptotic MCTPs.
Here, the asymptotic MCTP with $t$-distribution can be recommended for non-singular data with moderate sample sizes.
As the version with parametric bootstrap tends to be a bit conservative in comparison to the approach with wild bootstrap, this method can be recommended if someone is interested in more conservative test decisions or in the situation of moderate negative pairing.
Here the phenomena that lead to liberal and conservative behaviour counteract each other.

\subsection{Simulation Results Under the Alternative Hypothesis}
For further insights we also performed simulations under the alternative.
As the power is known to be higher for bigger sample sizes we consider in our power simulations the smaller sample sizes with multiplier $K\in\{1,2\}$.
And we consider only settings with $d=3$ dimensions, $k=3$ groups and Dunnett's testing problem.
To ensure that the data does not fulfil the null hypothesis we add $\delta\in\{0.5,1,1.5,2,3\}$ in three ways to $\boldsymbol{\mu}_i$, $i\in\{1,2,3\}$:
\begin{enumerate}
\item Shift alternative: $\boldsymbol{\mu}_1=\boldsymbol{\mu}_2=0,\,\boldsymbol{\mu}_3=\delta\cdot\mathbf{1}_{n_id}$,
\item One-point alternative: $\boldsymbol{\mu}_1=\boldsymbol{\mu}_2=0,\,\boldsymbol{\mu}_3=(\delta, 0,\dots,0)$,
\item Trend alternative: $\boldsymbol{\mu}_1=\boldsymbol{\mu}_2=0,\,\boldsymbol{\mu}_3=(\delta, \delta/2,\dots,\delta/d)$.
\end{enumerate}
\begin{figure}
\centering
\includegraphics[scale=0.7]{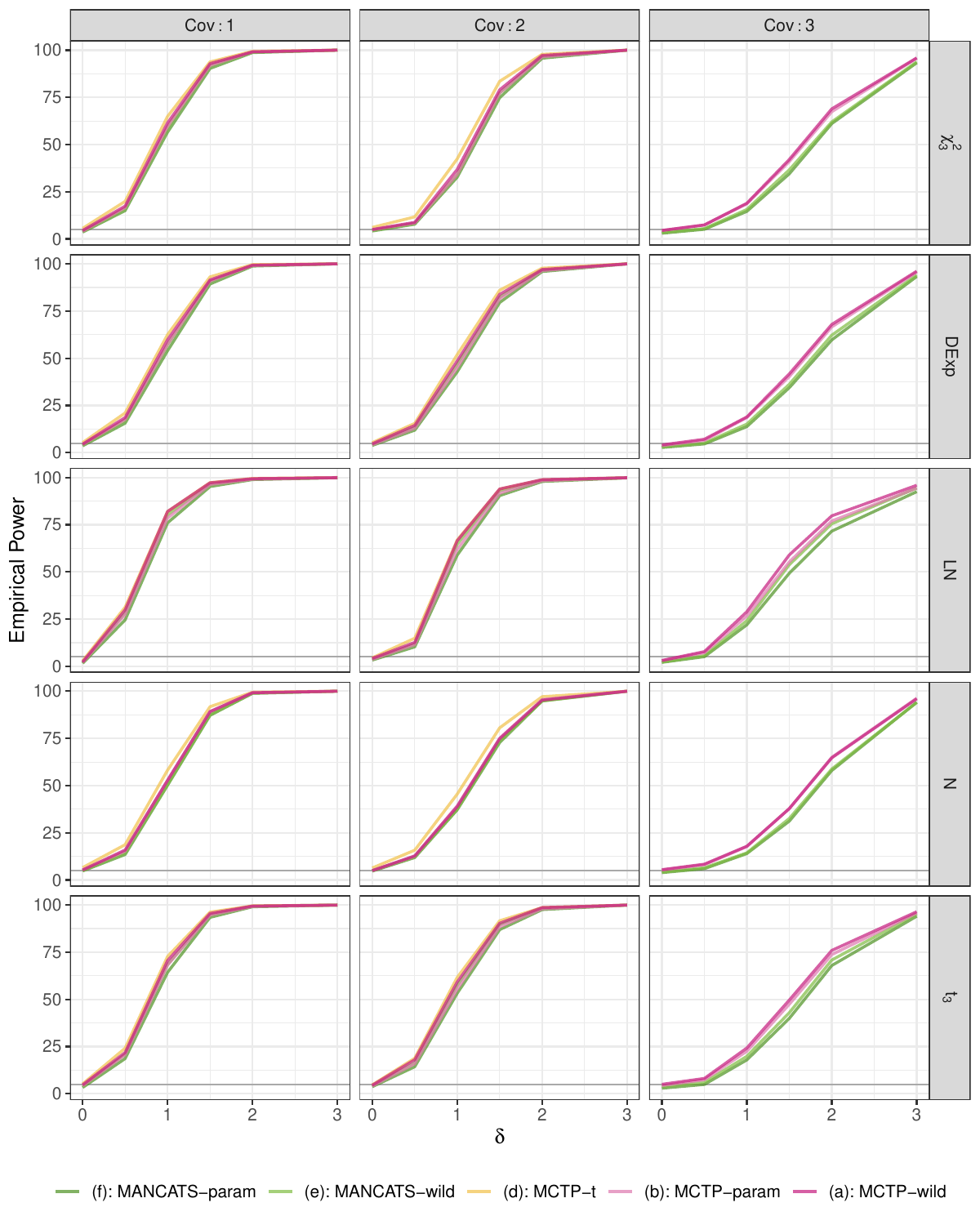}
\caption{Empirical power in $\%$ through a shift alternative for Dunnett's testing problem with $k=3$ and $d=3$, for the sample $1\cdot\,n_3^{(1)}=(10,10,10)$ and the covariance settings: homoscedastic (1), heteroscedastic (2) and singular (3), and distributions: standard normal distribution (\textsf{N}), t-distribution with $3$ degrees of freedom ($t_3$), $\chi^2$-distribution with $3$ degrees of freedom ($\chi_3^2$), logarithmic standard normal distribution (\textsf{LN}) and double exponential distribution (\textsf{DExp}).}
\label{fig:10.0_power_b}
\end{figure}
As the asymptotic MCTP with multivariate normal distribution has a liberal behaviour under the null hypothesis we exclude this method from our power analysis.
The results for this method can still be found in the Supplementary Material.

In Figure \ref{fig:10.0_power_b}, the power of the shift alternative of the 12 settings with the sample $1\cdot\,n_3^{(1)}=(10,10,10)$ is plotted.
Here, it can be observed that the five presented methods have a similar power.
For the singular covariance setting (3) the power is smaller in comparison to the other covariance settings for all methods and the resampling MCTPs have a better performance than the other methods.
For some heteroscedastic settings ($\chi^2_3$-and normal distribution) the MCTP with multivariate $t$-distribution has a little better power than the other methods.
This is not surprising because of the liberal behaviour of the tests in these settings.
As expected the power simulations with the multiplier $K=2$ show a faster increase of the power than for multiplier $K=1$, which is in line with the proven asymptotic behaviour of the methods.
In the Supplementary Material, there are power plots for all settings with sample factor $K=1$, especially for the different ways to create data under the alternatives.
As the one-point alternative is more difficult to detect for a statistical test, the power increases much slower for this alternative, especially for the singular covariance setting (3).
The increase of the power under the trend alternative is between the shift and the one-point alternative, which is not surprising.
In the settings with negative pairing (sample size $1\cdot\,n_3^{(2)}=(20,10,10)$, heteroscedastic covariance (2)), it is observable for all considered alternatives that the power increases very slowly, especially for $\delta=0.5$ the power is very small compared to other settings.
This phenomenon occurs in all considered methods, similarly to the bad performance for negative pairing regarding the FWER-control.
The power results for $K=2$ can be found in the Supplementary Material.
To conclude, when considering the power, there is no reason not to continue recommending the resampling MCTPs.
In general, these methods have a high power, even in singular settings.
Especially, the power is higher than the power of the Bonferroni-corrected MANCATs.

\section{Data Analysis}\label{sec:dataanalysis}

In this section, we illustrate the practical application of our MCTPs. 

\paragraph{Hypnosis interventional study.}
Our  analysis is motivated from a real interventional study, the \emph{HypnoTreat} study, 
conducted at Ulm University~\parencite{karrasch_effects_2022,karrasch_exploratory_2023,karrasch_randomized_2023}. 
As the data from the HypnoTreat is not published, we generated a synthetic dataset that preserves key characteristics of the HypnoTreat study, including
design, sample sizes, range, skewness, and dependency structures. 
The resulting dataset is published in \textcite{thiel_hypnotreatsynth_2025}.
HypnoTreat examined the effects of a single relaxation hypnosis session on several heart rate variability (HRV) parameters in chronically stressed individuals.
HRV refers to the variability in the interval between successive heartbeats (inter-beat intervals), reflecting the heart's capacity to respond to internal and external stimuli and maintain homeostasis under varying environmental demands \parencite{rajendra_acharya_heart_2006}.
HRV is influenced by both acute and chronic stress \parencite{kim_stress_2018}, and therefore, serves as a non-invasive marker of autonomic nervous system function and the dynamic interaction between its sympathetic and parasympathetic branches. 

In HypnoTreat, a total of 45 participants was randomly assigned to either an intervention group (listening to a 20-minute relaxation hypnosis via headphones) or a control group (watching a 20-minute documentary about the universe).
HRV was assessed using five standard time- and frequency-domain parameters, assessed at baseline and at post-intervention \parencite{sammito_nutzung_2012, kim_stress_2018, shaffer_overview_2017}:
\begin{description}
	\item[\textbf{SDNN}:] standard Deviation of Normal-to-Normal Intervals (measured in ms); reflects overall HRV and indicates both sympathetic and parasympathetic activity.
	\item[\textbf{RMSSD}:] root Mean Square of Successive Differences (measured in ms); primarily reflects short-term HRV and serves as a reliable marker of parasympathetic activity.
	\item[\textbf{HF}:] high Frequency (measured in ms$^2$); mainly associated with parasympathetic activity.
	\item[\textbf{LF}:] low Frequency (measured in ms$^2$); predominantly reflects sympathetic activity.
	\item[\textbf{VLF}:] very Low Frequency, (measured in ms$^2$); less well understood, but linked to other stress-regulating systems such as the endocrine and immune systems.
\end{description}

Besides these HRV parameters, two important confounding variables have been observed at baseline: \emph{suggestibility} and \emph{perceived chronic stress}.
Chronic stress was assessed using the German translation of the Perceived Stress Scale-14 \parencite{cohen_global_1983}, which is a sum of 4-point Likert scales and ranges from 0 to 56 (\textbf{PSS}).
Suggestibility, defined as an individual's responsiveness to hypnotic procedures, was measured using the German version of the Harvard Group Scale of Hypnotic Susceptibility (\textbf{HGSHA}) \parencite{bongartz_w_harvard_1982,shor_norms_1963}, where a total score was calculated based on the sum of responses to 12 items.

\paragraph{Data modelling.}

\begin{figure}[t]
\centering
\includegraphics[width=\textwidth]{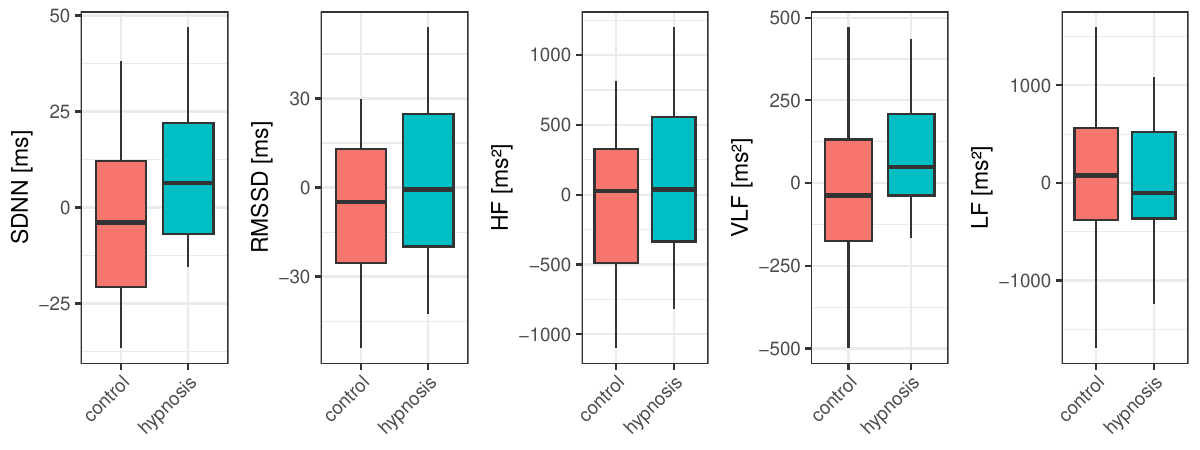}
\caption{Change from baseline of five heart rate variability (HRV) parameters.}
\label{fig:hrv-parameters}
\end{figure}

To analyse this longitudinal dataset, we first calculate the change from baseline of all HRV parameters, that is, we subtract the baseline value from the post-intervention values.
We now consider the newly gained change from baseline variables as the dependent variables of our analysis.
Note that using change from baseline is a simple way of correcting for baseline imbalances between subjects. 
\autoref{fig:hrv-parameters} displays the change from baseline for all HRV parameters in boxplots. 
Visual investigation suggests possible treatment effects in some variables. 
In the synthetic dataset, we introduced an artificial shift effect in the hypnosis group in two variables \textsf{SDNN} and \textsf{VLF}~\parencite{thiel_hypnotreatsynth_2025}.
Consequently, we are interested in whether the statistical inference tools proposed in this manuscript are able to detect these effects at a significance level of $\alpha = 0.05$, especially when controlling for multiplicity, which is clearly required for separate investigation of the five variables.
In fact, we are interested in an overall effect and in specific local effects, which leads to a multiple multivariate two-sample testing problem as explained in Section \ref{sec:mtestprob}. 
It can be realized by the hypothesis matrix $\mathbf{H}_d:=(1,-1)\otimes\mathbf{I}_5$.
Then, the global hypothesis is characterized as $\mathcal{H}^d_{0}:\,\mathbf{H}_d\boldsymbol{\mu}=\mathbf{0}\Leftrightarrow \boldsymbol{\mu}_{1}=\boldsymbol{\mu}_{2}$.
Moreover, the local hypotheses are defined by the rows of $\mathbf{H}_d$ and are consequently given by $\mathcal{H}^d_{0,\ell}:\mu_{1\ell}=\mu_{2,\ell}$, $\ell\in\{1,\dots,5\}$.

\begin{figure}[t]
\centering
\includegraphics[width=0.6\textwidth]{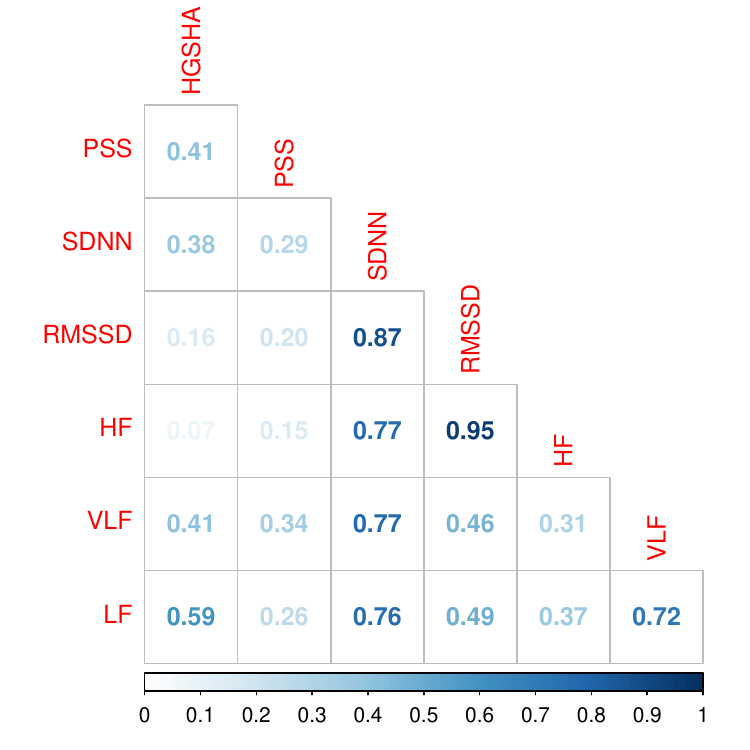}
\caption{(Pearson) correlation matrix of two baseline covariates (HGSHA, PSS) and change from baseline of five HRV parameters (SDNN, RMSSD, HF, LF, VLF).}
\label{fig:correlation}
\end{figure}

Before using an inferential tool, we inspect the correlation structure in the dataset in \autoref{fig:correlation}.
We observe a high correlation between the HRV parameters.
For instance, SDNN shows correlations of 0.76 -- 0.87 with the other HRV parameters, and the correlation between RMDSS and HF is even 0.95.
The high level of linear dependency reflected by these values suggests a multivariate modelling approach for the HRV parameters.
The subsequently presented approach offers such an advantage: when applied to our data, it models the joint distribution of test statistics for covariate-adjusted means of the HRV parameters (cf.\ Theorem \ref{thm:testdistribution}).
Importantly, the resampling MCTPs are also capable of leveraging information from the baseline covariates HGSHA and PSS.
\autoref{fig:correlation} shows small to medium correlation coefficients of HGSHA and PSS with several HRV parameters, which suggests that precision of group comparisons may be improved by controlling for these covariates.
For the considered methods, this is done in \autoref{equation:covariate-adjusted-mean}, where the mean estimators of HRV parameters are corrected for a linear dependency on the covariates.

\paragraph{Estimators.}

We consider covariate-adjusted means as estimands. 
With the notation from \autoref{sec:methods}, we have ${\hat{\boldsymbol{\mu}} = (\hat{\boldsymbol{\mu}}_1', \hat{\boldsymbol{\mu}}_2')'}$, where ${\hat{\boldsymbol{\mu}}_i = (\hat{\mu}_{i\ell})_{\ell = 1, \ldots 5}'}$ for treatment groups $i\in\{ 1, 2\}$.
Here, $\ell \in\{ 1 \dots, 5\}$ denotes the individual change from baseline HRV parameters.
Using the hypothesis matrix $\mathbf{H}_d$, we obtain the vector of covariate-adjusted mean differences $\hat{\boldsymbol{\mu}}_1-\hat{\boldsymbol{\mu}}_2$ between the two treatment groups.
\autoref{tab:mean-diff} displays the absolute values of these covariate-adjusted mean differences as well as unadjusted mean differences for comparison.
We observe that covariate adjustment leads to larger absolute mean differences.
Note that for the two parameters SDNN and VLF, for which we know that there exists a treatment effect, the absolute mean difference is even increased by approximately {1}/{3}.

\begin{table}[t]
\centering
\begin{tabular}{r|ccccc}
		& \textbf{SDNN} & \textbf{RMSSD} & \textbf{HF} & \textbf{VLF} & \textbf{LF} \\
	\hline
	unadjusted & 12.95 & 9.80 & 157.23 & 115.96 & 100.38\\
	\hline
	covariate-adjusted & 17.03 & 11.43 & 167.89 & 156.93 & 103.14\\
	\hline
\end{tabular}
\caption{\label{tab:mean-diff}Absolute value of unadjusted and covariate-adjusted mean differences of change from baseline HRV parameters between the two treatment groups.}
\end{table}

\paragraph{Adjusted significance levels \& p-values.}

We now apply the MCTPs proposed in \autoref{sec:methods} and \autoref{sec:simus}: (i) a wild and (ii) a parametric bootstrap MCTP, as well as (iii) a normally-distributed, and (iv) a $t$-distributed asymptotic MCTP as explained in Section \ref{sec:mtestprob}.
Moreover, we compare these MCTPs with (v) a wild and (vi) a parametric bootstrap MANCATS-based test.
For each of these tests, we compute local p-values for each contrast $\hat{\mu}_{1\ell} - \hat{\mu}_{2\ell}$, $\ell \in\{ 1, \dots, 5\}$.
Alongside with these local p-values, we compute adjusted significance levels $\gamma$ to compare all p-values.
For the bootstrap MCTPs, local p-values are defined in Equation (\ref{eq:local-p-value-bootstrap-mctp}) and $\gamma$ is defined in~Equation (\ref{eq:adjusted-level-bootstrap-mctp}), see Section \ref{sec:p-values} for the methodology.
For the asymptotic MCTPs, local p-values are defined in the Supplementary Material and $\gamma$ is obtained by plugging in the numerical value of the equicoordinate quantile into the (univariate) standard normal- or $t$-distribution function.
For the MANCATS-based tests, $\gamma$ is obtained by standard Bonferroni adjustment.

All approaches aim to control the global FWER $\alpha$ while providing coherent local and global test decisions.
As we have investigated in \autoref{sec:simus}, the individual methods differ in their capability of actually controlling $\alpha$.
Nevertheless, the representation in terms of local p-values and adjusted significance levels $\gamma$ guarantees comparability between the methods.
\autoref{tab:local-p-values} displays this representation for each method (i) -- (vi).
Here, we observe that the adjusted significance level $\gamma$ obtained from Bonferroni is with $0.0100$ smaller than all other adjusted levels.
The largest levels $\gamma$ can be observed with the asymptotic MCTPs (iii) and (iv), where the levels of the bootstrap MCTPs (i) and (ii) are only slightly below.
This represents the conservative behaviour of the Bonferroni-adjustment and the advantage of the MCTPs: they produce moderate adjusted levels while still controlling the FWER by leveraging dependencies between local test statistics.
Note that a more conservative behaviour of the Bonferroni-adjusted methods is expected in case of more than $5$ local hypotheses.
The order of the levels $\gamma$ displayed in \autoref{tab:local-p-values} are also in line with our simulation results from \autoref{sec:simus}, which attested (iii) and (iv) a somewhat liberal behaviour and (v) and (vi) a rather conservative behaviour.
Notably, all methods (i) -- (vi) reject at least one local hypothesis and 4 out of 6 methods correctly detect both effects in SDNN and in VLF.

\begin{table}[t]
	\centering
	\begin{tabular}{lrrrrrr}
	\toprule
		& $\boldsymbol{\gamma}$ & \textbf{SDNN} & \textbf{RMSSD} & \textbf{HF} & \textbf{VLF} & \textbf{LF} \\
		\midrule
		(i) MCTP-wild & 0.0145 & \textbf{0.0050} & 0.1270 & 0.3475 & \textbf{0.0110} & 0.5905\\
		
		(ii) MCTP-param & 0.0150 & 0.0155 & 0.2180 & 0.4390 & \textbf{0.0075} & 0.6275\\
		
		(iii) MCTP-norm & 0.0157 & \textbf{0.0025} & 0.1228 & 0.3603 & \textbf{0.0072} & 0.5923\\
		
		(iv) MCTP-t & 0.0162 & \textbf{0.0043} & 0.1302 & 0.3654 & \textbf{0.0102} & 0.5950\\
		
		(v) MANCATS-wild & 0.0100 & \textbf{0.0040} & 0.1180 & 0.3540 & \textbf{0.0080} & 0.6100\\
		
		(vi) MANCATS-param & 0.0100 & \textbf{0.0075} & 0.1445 & 0.3640 & 0.0115 & 0.5720\\
		\bottomrule
	\end{tabular}
	\caption{\label{tab:local-p-values}Local p-values for covariate-adjusted mean differences of five change from baseline HRV parameters and adjusted significance levels $\gamma$ for various testing methods. Significant test decisions are marked by p-values in bold. The number of bootstrap runs was set to 2000 and the significance level for the global hypothesis (= at least one local hypothesis is rejected) was set to $\alpha = 0.05$.}
\end{table}

\section{Discussion}\label{sec:end}
We extended the framework of multiple contrast testing procedures (MCTPs) to a general semiparametric MANCOVA model, enabling multivariate multiple comparisons based on covariate-adjusted means.
The proposed approach allows for heteroscedastic data, singular covariance matrices, and deviations from multivariate normality. 
This makes it particularly suited for complex interventional studies with multiple outcomes.
In this context, the main advantage of the model is the possibility to analyse several highly correlated outcomes together in one model.
Moreover, our model covers further multivariate testing problems like many-to-one and all-pair comparisons.

From a technical point of view, we considered a generalized calculation of critical values to comply with the flexible semiparametric model. 
This is different to traditional MCTPs that make use of equicorrdinate quantiles, and also necessitates resampling-based inference approaches. 
For the latter, we followed \textcite{friedrich_mats_2018} and \textcite{zimmermann_multivariate_2020}, and studied  parametric and wild bootstrap approaches. 
Both are theoretically justified within our model framework and yield favourable small-sample performance, as confirmed by extensive simulation studies.
In particular, our findings from the simulation study are in line with the theoretical results and demonstrate reliable type-I-error control and competitive power, even on small sample sizes and on singular data.
The illustrative data analysis, based on a synthetic dataset mimicking a psychological intervention study, highlights the practical relevance of the presented methods. 
It exemplifies how multivariate group comparisons can benefit from simultaneous covariate adjustment and multiplicity control.

For the future, several methodological extensions are promising: 
the recent preprint by \textcite{sattler_quadratic_2024} provides multivariate MCTPs based on quadratic form type statistics instead of linear ones, which allows multiple testing on a vector of group-wise means. 
This procedure could also be adapted for covariate-adjusted vectors of means.
Thinking ahead, covariate-adjusted multiple testing of other estimands beyond the mean, e.g. based on quantiles or nonparametric relative effect, are of high interest:
in particular, an extension of the quantile based (M)ANOVA approaches of \textcite{ditzhaus_qanova_2021, baumeister_quantile-based_2024} to allow for covariate-adjustments is tempting. 
Similarly, a combination of covariate-adjustments \parencite{bathke_nonparametric_2003, thiel_resampling_2024} and MCTPs \parencite{konietschke_rank-based_2012,noguchi_nonparametric_2020} for nonparametric relative effects would be interesting.
Both, quantile- and rank-based approaches, could be an opportunity to overcome the bad performance on skewed data, which was observed in the simulation study for log-normal data.
An \textsf{R} implementation of our proposed MCTPs and those by \textcite{zimmermann_multivariate_2020} is currently in preparation. 
Until then, the software implementation we used for the data analysis is provided in the Supplementary Material.

\newpage
\section*{CRediT Authorship Contribution Statement}

\textbf{Marléne Baumeister:} Conceptualization, Data Curation, Formal Analysis, Methodology, Project Administration, Software, Visualization, Writing – Original Draft Preparation;
\textbf{Konstantin Emil Thiel:} Data Curation, Formal Analysis, Methodology, Project Administration, Software, Visualization, Writing – Original Draft Preparation;
\textbf{Lynn Matits:} Data Curation, Investigation, Resources, Writing – Original Draft Preparation;
\textbf{Markus Pauly:} Conceptualization, Funding Acquisition, Methodology, Supervision, Writing – Review \& Editing;
\textbf{Georg Zimmermann:} Conceptualization, Writing – Review \& Editing;
\textbf{Paavo Sattler:} Formal Analysis, Methodology, Supervision, Writing – Review \& Editing.

\section*{Acknowledgement}

MB and MP has been partly supported by the Research Center Trustworthy Data Science and Security (\url{https://rc-trust.ai}), one of the Research Alliance centers within the \href{https://uaruhr.de}{UA Ruhr}.
The authors gratefully acknowledge the computing time provided on the Linux HPC cluster at TU Dortmund University (LiDO3), partially funded in the course of the Large-Scale Equipment Initiative by the Deutsche Forschungsgemeinschaft (DFG, German Research Foundation) as project 271512359.
LM was supported by a Ph.D. scholarship from the German Academic Scholarship Foundation (Studienstiftung des deutschen Volkes).

We thank Merle Munko for the fruitful exchange about the methodology of multiple contrast tests and our student assistant Mitja Schemmer for the help with the programming of the simulations.

\section*{Declarations of interest: none}

\section*{Data Availability Statement}
The data that support the findings of this work e.g. all simulation results, simulation scripts and the scripts for the data analysis are openly available in TUDOdata at \url{https://doi.org/10.17877/TUDODATA-2025-MANOBADL}.


\printbibliography	
	
\newpage
\appendix
\section{Proofs} \label{sec:proofs}

\subsection{Proof of Theorem \ref{thm:testdistribution}}
To derive the asymptotics of $\mathbf{A}_n(\mathbf{H})$ and $A_n(\mathbf{h}_s),\,s\in\{1,\dots,r\},$ we can deduce the following statements from Proposition \ref{thm:clt}. 
To extract the vectors $\boldsymbol{\mu}$ and $\hat{\boldsymbol{\mu}}$ from $\boldsymbol{\beta}$ and $\hat{\boldsymbol{\beta}}$ we use the vector $\tilde{\mathbf{h}}'_s=(\mathbf{h}'_s,\mathbf{0}'_c)'\in\br^{dk + c}$ and the zero-inflated covariance estimator
\begin{align*}
\hat{\tilde{\mathbf{D}}} = \hat{\mathbf{D}} \oplus \mathbf{0}_{c,c}
\end{align*} 
for all $s\in\{1,\dots,r\}$.
The matrix $\tilde{\mathbf{D}}$ is obtained from $\mathbf{D}$ in the same way.
The zero-inflated contrast matrix $\tilde{\mathbf{H}}=(\tilde{\mathbf{h}}_1,\dots,\tilde{\mathbf{h}}_r)'\in\br^{r\times (dk + c)}$ is consequently built with the zero-inflated contrast vectors.
Then, it holds $\mathcal{H}_{0,s}:\,\tilde{\mathbf{h}}_s\boldsymbol{\beta}=\mathbf{h}_s\boldsymbol{\mu}=\mathbf{0}$, $\mathcal{H}_{0}:\,\tilde{\mathbf{H}}\boldsymbol{\beta}=\mathbf{H}\boldsymbol{\mu}=\mathbf{0}$ and $\tilde{\mathbf{h}}'_s\tilde{\mathbf{D}}\tilde{\mathbf{h}}_s=\mathbf{h}'_s\hat{\mathbf{D}}\mathbf{h}_s$.
Under $\mathcal{H}_{0}$, by Slutzky's and the Continuous Mapping Theorem we can describe the distribution of the vector of test statistics:
\begin{align*}
A_n(\mathbf{H})&=(\mathbf{H}\hat{\mathbf{D}}\mathbf{H}')_0^{-1/2}\sqrt{n}\mathbf{H}\hat{\boldsymbol{\mu}}
=(\tilde{\mathbf{H}}\hat{\tilde{\mathbf{D}}}\tilde{\mathbf{H}}')_0^{-1/2}\sqrt{n}\tilde{\mathbf{H}}\hat{\boldsymbol{\beta}}\\
&=(\tilde{\mathbf{H}}\hat{\tilde{\mathbf{D}}}\tilde{\mathbf{H}}')_0^{-1/2}\tilde{\mathbf{H}}\sqrt{n}\left(\hat{\boldsymbol{\beta}}-\boldsymbol{\beta}\right)
\stackrel{d}{\longrightarrow}(\tilde{\mathbf{H}}\tilde{\mathbf{D}}\tilde{\mathbf{H}}')_0^{-1/2}\tilde{\mathbf{H}}\mathbf{Z}
=:\mathbf{B},
\end{align*}
From the distribution of $\mathbf{Z}$ it follows that $\mathbf{B}=(B_1,\dots,B_r)$ has a multivariate normal distribution with expectation $\E(\mathbf{A}_n(\mathbf{H}))=\mathbf{0}$ and covariance matrix 
\begin{align*}
\mathbf{R}:=\cov\left(\mathbf{A}_n(\mathbf{H})\right)=\left(\mathbf{H}\mathbf{D}\mathbf{H}'\right)_0^{-\frac{1}{2}}\mathbf{H}\boldsymbol{\Lambda}_{11}\mathbf{H}'\left(\mathbf{H}\mathbf{D}\mathbf{H}'\right)_0^{-\frac{1}{2}}.
\end{align*}
This is the first assertion.
Furthermore, $A_n(\mathbf{h}_s)$ is for every $s\in\{1,\dots,r\}$ already under $\mathcal{H}_{0,s}$ asymptotically distributed like $B_s$.
In general, we can argue that $|A_n(\mathbf{h}_s)|/\sqrt{n}$ converges in probability:
\begin{align*}
\frac{1}{\sqrt{n}}\left|A_n(\mathbf{h}_s)\right|= \frac{\left|\mathbf{h}'_s\hat{\boldsymbol{\mu}}\right|}{\sqrt{\mathbf{h}'_s\hat{\mathbf{D}}\mathbf{h}_s}}\stackrel{p}{\longrightarrow}\frac{\left|\mathbf{h}'_s\boldsymbol{\mu}\right|}{\left|\sqrt{\mathbf{h}'_s\mathbf{D}\mathbf{h}_s}\right|}.
\end{align*}
Under $\mathcal{H}_{1,s}$ this limiting value is greater than zero.
From this we can conclude that under $\mathcal{H}_{1,s}$ the test statistics $A_n(\mathbf{h}_s)$ converges in probability to $\infty$ for all $s\in\{1,\dots,r\}$, which is the second assertion.

\subsection{Proof of Theorem \ref{thm:wildteststatistic}}
In \textcite{zimmermann_multivariate_2020} statement (C1) in the proof of Theorem 3 is a Central Limit Theorem for the wild bootstrap:
\begin{align*}
\sqrt{n}\hat{\boldsymbol{\beta}}^*\stackrel{d}{\longrightarrow}\mathbf{N}\sim\mathcal{N}\left(\mathbf{0},\boldsymbol{\Lambda}\right)\quad\text{given the data}.
\end{align*}
From this and from the consistency of $\hat{\mathbf{D}}^*$ \parencite[Proof of Theorem 3]{zimmermann_multivariate_2020} it follows similarly to the original test statistic:
\begin{align*}
A^*_n(\mathbf{h}_s)=\sqrt{n}\frac{\mathbf{h}_s'\hat{\boldsymbol{\mu}}^*}{\sqrt{\mathbf{h}_s'\hat{\mathbf{D}}^*\mathbf{h}_s}}
=\tilde{\mathbf{h}}'_s\frac{\sqrt{n}\hat{\boldsymbol{\beta}}^*}{\sqrt{\tilde{\mathbf{h}}'_s\hat{\tilde{\mathbf{D}}}^*\tilde{\mathbf{h}}_s}}
\stackrel{d}{\longrightarrow}\tilde{\mathbf{h}}'_s\frac{\mathbf{N}}{\sqrt{\tilde{\mathbf{h}}'_s\tilde{\mathbf{D}}\tilde{\mathbf{h}}_s}}
=\mathbf{B}.
\end{align*}
This is close to the assertion.

\subsection{Proof of Theorem \ref{thm:parateststatistic}}
In the Proof of Theorem 4 in \textcite{zimmermann_multivariate_2020} there is the following statement given the data:
\begin{align*}
\sqrt{n}\hat{\boldsymbol{\beta}}^{\star}\sim\mathcal{N}\left(\mathbf{0},\hat{\boldsymbol{\Lambda}}_n\right),
\end{align*}
where $\hat{\boldsymbol{\Lambda}}_n=\cov(\sqrt{n}\hat{\boldsymbol{\beta}}^{\star}\vert\mathbf{Y})$ is the conditional covariance estimator of $\hat{\boldsymbol{\beta}}^{\star}$, which converges in probability to $\boldsymbol{\Lambda}$.
From this it follows the assertion.

\subsection{Proof of Theorem \ref{thm:adjlevel}}
It is stated in the proof of Theorem 6 in \textcite{munko_rmst-based_2024}, that Lemma S5 and S6 in the Supplement of \textcite{munko_rmst-based_2024} imply $q^{\circ}_{s,1-\gamma_n(\alpha)}\stackrel{P}{\longrightarrow}q_{s,\text{FWER}_n^{-1}(\alpha)},\, s\in\{1,\dots,r\}$.
The two Lemmas are applicable because of the identical framework in \textcite{munko_rmst-based_2024}. 
Let $T\subset\{1,\dots,r\}$ be the set of true hypotheses.
By Slutzky's Theorem and Equation (\ref{equ:testdistr}) it follows
\begin{align*}
\left(A_n(\mathbf{h}_s),q^{\circ}_{s,1-\gamma_n(\alpha)}\right)_{s\in T}\stackrel{d}{\longrightarrow}\left(B,q_{s,\text{FWER}_n^{-1}(\alpha)}\right)_{s\in T}.
\end{align*}
Analogously to the proof of Theorem 6 in \textcite{munko_rmst-based_2024} we show that
\begin{align}
\nonumber\text{Pr}\left(\max_{s\in T}A_n(\mathbf{h}_s)>q^{\circ}_{s,1-\gamma_n(\alpha)}\right)&=1-\text{Pr}\left(\forall s\in T:A_n(\mathbf{h}_s)\le q^{\circ}_{s,1-\gamma_n(\alpha)}\right)\\
\nonumber &\longrightarrow 1-\text{Pr}\left(\forall s\in T:B\le q_{s,\text{FWER}_n^{-1}(\alpha)}\right)\\
&\le 1-\text{Pr}\left(\forall s\in \{1,\dots,r\}:B\le q_{s,\text{FWER}_n^{-1}(\alpha)}\right)\label{equ:equal}\\
\nonumber &=\text{FWER}_n\left(\text{FWER}^{-1}_n(\alpha)\right)=\alpha.
\end{align}
Note that there is an equality in (\ref{equ:equal}) if $T=\{1,\dots,r\}$.

\subsection{Proof of Proposition \ref{thm:p-value-equivalence}}
For (i), let $s\in\{1,\dots,r\}$ be fixed.
Firstly, assume $p_{n,s}\le\gamma_n(\alpha)$ and for a proof by contradiction $\varphi^{\circ}_{n,s}=0$, which is equivalent to $\vert A_n(\mathbf{h}_s)\vert\le q^{\circ}_{s,1-\gamma_n(\alpha)}$.
Then,
\begin{align*}
p_{n, s} = \frac{1}{B} \sum_{b = 1}^{B} \mathds{1} \left\{ | A^{\circ,b}_n(\mathbf{h}_s) | \geq | A_n(\mathbf{h}_s) | \right\}
\ge \frac{1}{B} \sum_{b = 1}^{B} \mathds{1} \left\{ | A^{\circ,b}_n(\mathbf{h}_s) | \geq q^{\circ}_{s,1-\gamma_n(\alpha)} \right\}=\gamma_n(\alpha)+\frac{1}{B},
\end{align*}
where the last equality holds by the definition of the quantile $q^{\circ}_{s,1-\gamma_n(\alpha)}$.
From this, it follows the contradiction $\gamma_n(\alpha)\ge\gamma_n(\alpha)+1/B$.
Secondly, we assume $\varphi^{\circ}_{n,s}=1$.
Then, $\vert A_n(\mathbf{h}_s)\vert> q^{\circ}_{s,1-\gamma_n(\alpha)}$.
From this, with the definition of the quantile $q^{\circ}_{s,1-\gamma_n(\alpha)}$, it can be concluded
\begin{align*}
p_{n, s} = \frac{1}{B} \sum_{b = 1}^{B} \mathds{1} \left\{ | A^{\circ,b}_n(\mathbf{h}_s) | \geq | A_n(\mathbf{h}_s) | \right\}
\le \frac{1}{B} \sum_{b = 1}^{B} \mathds{1} \left\{ | A^{\circ,b}_n(\mathbf{h}_s) | \geq q^{\circ}_{s,1-\gamma_n(\alpha)} \right\}
\le\gamma_n(\alpha).
\end{align*}
To prove (ii), we argue that $p_n=\min\{p_{n,1},\dots,p_{n,r}\}\le\gamma_n(\alpha)$ if and only if there exist $s\in\{1,\dots,r\}$ such that $p_{n,s}\le\gamma_n(\alpha)$.
Due to (i), this is equivalent to the situation that there exist $s\in\{1,\dots,r\}$ with $\varphi^{\circ}_{n,s}=1$, which is the same as $\varphi_n^{\circ}=1$.
Compare also Proposition 1 in \textcite{munko_rmst-based_2024}.
\end{document}